\documentclass[aps,pre,twocolumn,superscriptaddress,a4paper,showpacs,floatfix]{revtex4}
\usepackage{graphicx,psfig,amsmath,amssymb,epic,eepic,subeqnarray}
\usepackage{natbib}
\usepackage{easymat}
\usepackage{easybmat}
\usepackage{enumerate}

\newcommand{\lapv}{\Delta}
\newcommand{\lap}{\nabla^2}
\newcommand{\grad}{\nabla}
\newcommand{\bU}{{\bf U}}

\newcommand{\bV}{{\bf V}}
\newcommand{\bv}{{\bf v}}
\newcommand{\pd}{\partial}
\newcommand{\dt}{\Delta t}
\newcommand{\be}{{\bf e}}
\newcommand{\ml}{\mathcal{L}}
\def \RacRezero {2260}      
\def \RacRezeroPFtwo {6640} %
\def \RacRezeroSN {28\,445} %
\def \RacReforty {3\,125}   
\def \RacReninety {8\,622}  
\def \RaTakens {11\,750}    
\def \RetranstriD {54.8}    
\begin{document}

\title{
Influence of counter-rotating von K\'arm\'an flow \\
on cylindrical Rayleigh-B\'enard convection}
\author{Lyes Bordja}
\affiliation{Facult\'e Sciences de l'Ing\'enieur, D\'ept. G\'enie M\'ecanique
Univ. de Jijel, BP 98, Ouled Aissa, Jijel-18000, Algeria}
\author{Laurette S. Tuckerman}
\affiliation{PMMH-ESPCI, CNRS (UMR 7636), Univ. Paris VI \& VII, 
10 rue Vauquelin, 75231 Paris France}
\email[]{laurette@pmmh.espci.fr}
\homepage[]{http://www.pmmh.espci.fr/~laurette}
\author{Laurent Martin Witkowski}
\affiliation{LIMSI (UPR 3251) CNRS, Univ.~Paris VI, BP 133, 91403 Orsay, France}
\author{Mar\'ia Cruz Navarro}
\affiliation{Dept. Matem\'aticas, Facultad de CC.~Qu\'{\i}micas,
Univ. Castilla-La Mancha, 13071 Ciudad Real, Spain}
\author{Dwight Barkley}
\affiliation{Mathematics Institute, University of Warwick, 
Coventry CV4 7AL, United Kingdom}
\author{Rachid Bessaih}
\affiliation{LEAP, 
D\'ept. G\'enie M\'ecanique, 
Univ. Mentouri-Constantine, 
Route d'Ain El. Bey, 25000 
Constantine, Algeria}


\begin{abstract}
The axisymmetric flow in an aspect-ratio-one cylinder whose upper and lower
bounding disks are maintained at different temperatures and rotate at equal
and opposite velocities is investigated.  In this combined
Rayleigh-B\'enard/von K\'arm\'an problem, the imposed temperature gradient is
measured by the Rayleigh number $Ra$ and the angular velocity by the Reynolds
number $Re$.  Although fluid motion is present as soon as $Re\neq 0$, a
symmetry-breaking transition analogous to the onset of convection takes place
at a finite Rayleigh number higher than that for $Re=0$.  For
$Re<95$, the transition is a pitchfork bifurcation to a pair of steady states,
while for $Re>95$, it is a Hopf bifurcation to a limit cycle.  The steady
states and limit cycle are connected via a pair of SNIPER bifurcations except
very near the Takens-Bogdanov codimension-two point, where the scenario
includes global bifurcations. Detailed phase portraits and bifurcation
diagrams are presented, as well as the evolution of the leading part of the
spectrum, over the parameter ranges $0\leq Re\leq 120$ and $0\leq Ra \leq
30\,000$.
\end{abstract}

\date{\today}
\maketitle

\section{Introduction}

Thermal convection and shear are central to the study of hydrodynamic
instabilities. The interest in these stems both from their large number of
practical applications and also from their status as prototypes in theoretical
investigation.  In this study, we combine two well-known cylindrical
configurations: Rayleigh-B\'enard convection in which the upper and lower
bounding disks of the cylinder are maintained at different temperatures, and
von K\'arm\'an flow in which the disks rotate at equal and opposite
velocities.  We choose the simplest small-aspect-ratio geometry: a cylinder
with equal height and radius, with imposed axisymmetry.

There is an extensive literature on Rayleigh-B\'enard convection in a small to
medium aspect-ratio cylinder (in which the radius is between 0.5 and 5 times
the height),
e.g.~\cite{ChaSan71,BueCat,WanKuhRat,Behringer,TouHadHen,Boronska_JFM}.  The
references most relevant to us are axisymmetric simulations
\cite{Tuck_Bark_88,Tuck_Bark_89,Siggers} which produced patterns
of radially travelling concentric rolls (target patterns).  These were shown
to be produced in most cases by a saddle-node infinite period
(SNIPER) bifurcation which will be the subject of much of the present 
investigation.

The literature on von K\'arm\'an flow is not as voluminous, but is growing.
Restricting ourselves to the small to medium aspect-ratio and the exactly or
nearly exactly counter-rotating configuration, recent numerical articles
include axisymmetric and non-axisymmetric simulations 
\cite{GBYS_96,L_98,LHMKS_02,MGL_03,Nore_03,Nore_04,Nore_LMW_06}.  Interest in this
problem has been enhanced by experiments using this configuration at
high Reynolds numbers to generate turbulence
\cite{Brachet,Ravelet_04} or a magnetic field via the dynamo
effect \cite{Berhanu_07,Ravelet_PRL_08,Petrelis_09}.

The combined Rayleigh-B\'enard/von K\'arm\'an configuration seems, however,
unexplored.  This situation stands in contrast to that of other shear and
rotating flows.  The superposition of plane Poiseuille flow with
Rayleigh-B\'enard convection (PRB), has been the subject of a great many
investigations; indeed, a recent review \cite{Nicolas} cites hundreds of
articles.  For longitudinal rolls in plane Poiseuille and Couette flows, the
threshold remains unchanged \cite{CB_91,CB_92}.  For rolls transverse to the
Poiseuille flow, the threshold increases and its nature changes from absolute
to convective, e.g.~\cite{Muller_92}.  In Taylor-Couette flow between
cylinders at different temperatures, the nature of the instability also
depends strongly on the angle between the imposed shear and buoyancy force,
e.g.~\cite{TaggWeidman,Lepiller}. Turning to rotating disks, heat transfer in
the rotor-stator configuration (between a rotating and a stationary disk) is
extensively studied because of its applications to turbomachinery,
e.g. \cite{OwenRogers1}.  Rotating Rayleigh-B\'enard convection has long
interested geophysicists, as well as researchers in pattern formation
e.g.~\cite{Knobloch_98}, because of the possibility of chaos at onset; SNIPER
bifurcations are found in this system as well \cite{LRM_06,RLM_08}.  Of the
studies we have found of the flow between counter-rotating disks maintained at
different temperatures, one computes the similarity solution between disks of
infinite radius prior to the onset of convection \cite{Soong}, and another
considers turbulent flows and neglects buoyancy \cite{Hill_Ball}.

The Rayleigh-B\'enard/von K\'arm\'an problem in this geometry 
turns out to display a number of fascinating phenomena.
For this reason, this problem has been used to develop and 
test a reduced model for numerical simulation, in which fields 
are expanded in terms of the eigenfunctions of the governing equations 
linearized about a steady state; this study is 
the subject of a companion paper \cite{Maria_09}.

Our paper is organized as follows.
Section \ref{sec:problem} presents the governing equations of our
configuration, along with the non-dimensionalizations we have used.  In
section \ref{sec:methods}, we describe the various codes we have used to carry
out time-dependent simulations, branch continuation and linear stability
analysis. Section \ref{sec:thresholds} shows how the transition threshold is
affected by the disk counter-rotation.  In sections \ref{sec:steady} and
\ref{sec:oscill}, we present the steady states and limit cycles resulting from
the transition. The connection between the steady states and limit cycle 
via a SNIPER bifurcation is explained in section \ref{sec:bifs}, 
which presents a complete phase diagram and bifurcation analysis. Reversing the
traditional order, section \ref{sec:linear} gives a linear
stability analysis of the basic state, showing how the convective eigenvalues
and eigenvectors are interlaced and joined by the von K\'arm\'an flow.

\section{Problem formulation}
\label{sec:problem}

\begin{figure}[h]
\centerline{
\psfig{file=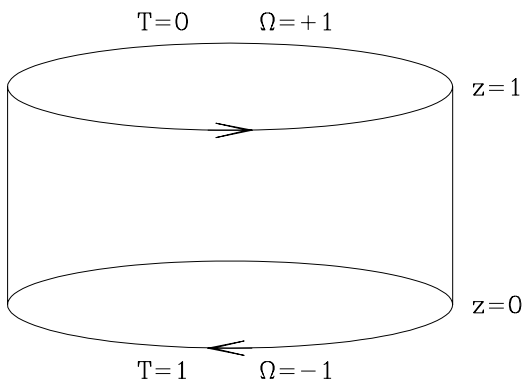,width=8cm}}
\caption{Cylinder with height and radius equal to one.  The bottom disk is
  heated and rotated in the clockwise direction, while the top disk is
  cooled and rotated in the counter-clockwise direction.}
\label{fig:schematic}
\end{figure}

We consider a viscous Newtonian fluid governed by the Boussinesq approximation
and contained in a cylinder of radius $R$ and height $H$.
The velocity and temperature fields are assumed axisymmetric. The bounding
cylinder is stationary and thermally insulating.  The upper and lower bounding
disks are thermally conducting, rotate with angular velocities $+\Omega$ and
$-\Omega$ and are held at temperatures $T_0$ and $T_0+\Delta T$, respectively.
The nondimensional parameters are defined as:
\begin{equation}
Re = \frac{\Omega R^2}{\nu},
\; Ra = \frac{g \gamma \Delta T R^3}{\alpha \nu},
\; Pr = \frac{\nu}{\alpha},
\; \Gamma = \frac{R}{H}
\end{equation}
where $\nu$, $\alpha$, $\gamma$ and $g$ are the kinematic viscosity, the thermal
diffusivity, the thermal expansion coefficient, and the gravitational
acceleration, respectively. 
The Prandtl number $Pr$ and aspect ratio $\Gamma$ are set
to one.  Reynolds numbers in the range $0\leq Re\leq 120$ will be studied,
along with Rayleigh numbers in the range $0\leq Ra\leq 30000$.

Our study uses several independent codes, summarized in section 
\ref{sec:methods}, which use different
non-dimensionalizations, and different problem formulations, i.e.  primitive
variables or streamfunction-vorticity.  To facilitate later discussion, we
briefly present these various formulations here.  The non-dimensionalized
temperature is taken to be the deviation from $T_0$ divided by $\Delta T$ and
lengths are non-dimensionalized by $R$.  The imposed counter-rotation of the
disks provides a natural timescale: if time is non-dimensionalized by
$1/\Omega$ and velocities by $R\Omega$, the resulting governing equations are:
\begin{subequations}
\begin{eqnarray}
\grad \cdot \bU &=& 0 \\
\pd_t \bU + (\bU\cdot \grad) \bU &=& -\grad P + \frac{1}{Re} \lapv \bU 
+\frac{Ra}{Pr Re^2} T \be_z 
\label{eq:evol_u}\\
\pd_t T + \bU \cdot \grad T &=& \frac{1}{Pr Re} \lap T
\label{eq:evol_T}
\end{eqnarray}
\label{eq:goveqns}
\end{subequations}
where $\lapv$ and $\lap$ denote the vector and scalar Laplacian, respectively.
The boundary conditions are:
\begin{subequations}
\begin{eqnarray}
U_r =  U_z = 0 ,\; U_\theta = +r, \; T = 0 
\;\mbox{ at }\; z=1 \\
U_r =  U_z = 0 ,\; U_\theta = -r ,\; T = 1 
\;\mbox{ at }\; z=0 \\
U_r = U_z = 0 ,\; U_\theta= 0,\; \pd_r T = 0 
\;\mbox{ at }\; r=1 \\
U_r = \pd_r U_z = 0 ,\; U_\theta= 0,\; \pd_r T = 0 
\;\mbox{ at }\; r=0
\end{eqnarray}
\label{eq:bcs}
\end{subequations}
These equations cannot be used in the absence of rotation, when $Re=0$.  To
allow for this case, the unit of time can instead be taken to be the viscous
diffusion time $R^2/\nu$, with the resulting velocity scale $\nu/R$.
%
In addition, 
with the assumption of axisymmetry, the meridional velocity components 
$(V_r,V_z)$ are best described by the Stokes streamfunction $\Psi$:
\begin{subequations}
\label{eq:strmvort}
\begin{equation}
V_r \be_r + V_z \be_z =\frac{\be_r}{r}\pd_z\Psi-\frac{\be_z}{r}\pd_r\Psi
=\frac{\be_\theta}{r} \times \grad\Psi
\end{equation}
Defining operators
\begin{equation}
\pd_\pm \equiv \pd_r \pm \frac{1}{r}, 
\end{equation}
the azimuthal vorticity is:
\begin{equation}
D^2\Psi \equiv 
\frac{1}{r}\left(\lap -\frac{2}{r}\partial_r\right)\Psi 
= \frac{1}{r}\left(\pd_-\pd_r + \pd_z^2\right)\Psi\label{eq:psi}
\end{equation}
\end{subequations}
and the Navier-Stokes and Boussinesq equations reduce to:
\begin{subequations}
\label{eq:eqstrm}
\begin{eqnarray} 
\partial_{s} D^2\Psi + \left(V_r\pd_-+V_z\pd_z\right) \:D^2\Psi
 & = & \left(\lap-\frac{1}{r^2}\right)D^2\Psi\nonumber \\
-\frac{Ra}{ Pr} \partial_r T 
&+& \partial_z   \left( \frac{V_\theta^2}{r}\right)\label{eq:omega}\\
\partial_{s} T +  (V_r\pd_r+V_z\pd_z) \:T &=& \frac{1}{Pr} \lap T \label{eq:tmp}\\
\partial_{s} V_\theta + \left(V_r\pd_++V_z\pd_z\right)  \:V_\theta
&=& \left(\lap - \frac{1}{r^2}\right) V_\theta \label{eq:vtheta}
\end{eqnarray}
\end{subequations}
The rotation then appears in the boundary conditions, which become:
\begin{subequations}
\begin{eqnarray}
\Psi = \partial_z \Psi = 0 \Longleftrightarrow V_r =  V_z = 0, \nonumber\\ 
V_\theta = +r Re ,\; T = 0 \mbox{ at } z=1 \label{eq:bcsz1}\\
\Psi = \partial_z \Psi = 0 \Longleftrightarrow V_r =  V_z = 0, \nonumber\\ 
V_\theta = -r Re ,\; T = 1 \mbox{ at } z=0 \label{eq:bcsz0}\\
\Psi = \partial_r \Psi = 0 \Longleftrightarrow V_r = V_z = 0,\nonumber\\ 
V_\theta=0,\; \pd_r T = 0 \mbox{ at } r=1 \\
\Psi = D^2\Psi = 0 \Longleftrightarrow V_r = \pd_r V_z = 0,\nonumber\\ 
V_\theta=0,\; \pd_r T = 0 \mbox{ at } r=0
\end{eqnarray}
\label{eq:bcstrm}
\end{subequations}
The velocity and time in \eqref{eq:goveqns}-\eqref{eq:bcs} are related
to those in \eqref{eq:eqstrm}-\eqref{eq:bcstrm}
by $\bV=Re\,\bU$ and $s=t/Re$;
the temperature $T$ and all lengths remain unchanged. 
Equations (\ref{eq:vtheta}) and (\ref{eq:tmp}) can be considered to be 
advection-diffusion equations for the azimuthal velocity $V_\theta$ 
and the temperature $T$. Equation (\ref{eq:omega}) shows that gradients 
in $V_\theta$ and $T$ in turn generate vorticity. 
The vertical gradient $\pd_z({V_\theta}^2/r)$, when combined with the
boundary conditions (\ref{eq:bcsz1})-(\ref{eq:bcsz0}), corresponds to Ekman
pumping, by which the $z$-dependent $V_\theta$ generates meridional vorticity
and velocity for any non-zero $Re$.  This meridional velocity in turn affects
the azimuthal velocity and the temperature via equations \eqref{eq:vtheta} and
\eqref{eq:tmp}.''
However, the thermal gradient term $\pd_r T$ in the vorticity equation 
(\ref{eq:omega}) is radial, and so the axial temperature gradient imposed by 
boundary conditions in (\ref{eq:bcsz1})-(\ref{eq:bcsz0})
does not immediately impart motion to the fluid for any non-zero $Ra$.
That is, Rayleigh-B\'enard convection occurs via an instability 
at a finite threshold $Ra_c$, in contrast to natural convection,
in which an externally imposed horizontal gradient generates motion for any 
$Ra>0$ via the term $Ra \,\pd_r T$.

Finally, we state the symmetries of the configuration of our problem,
which can be seen from figure \ref{fig:schematic}.
The Rayleigh-B\'enard convection problem has the symmetry of 
combined reflection in $z$ and in $T$ about their mean values,
sometimes called the Boussinesq symmetry.
The von K\'arm\'an flow has the symmetry of combined reflection 
in $z$ and in $\theta$, called $R_\pi$ \cite{Nore_04}.
(In this axisymmetric problem, reflection in $\theta$ means only 
reversing the sign of $U_\theta$; since all solutions are invariant under
rotations in $\theta$, these are not considered.)
The Rayleigh-B\'enard/von K\'arm\'an problem thus has the symmetry
which combines reflection in $z$, $\theta$ and $T$. 
A state is reflection-symmetric if it is invariant under
the reflection operator defined in the two formulations by:
\begin{eqnarray}
\kappa:\left(\begin{array}{c} U_r \\ U_\theta \\ U_z \\ T \end{array} \right)
(r,z) \equiv
\left(\begin{array}{r} U_r \\ -U_\theta \\ -U_z \\ 1-T \end{array} \right)
(r,1-z)
\quad\mbox{or}\nonumber\\
\kappa:\left(\begin{array}{c} \Psi \\ U_\theta \\ T \end{array} \right)
(r,z) \equiv
\left(\begin{array}{r} -\Psi \\ -U_\theta \\ 1-T \end{array} \right)
(r,1-z)
\label{eq:reflection}\end{eqnarray}

\section{Numerical Methods}
\label{sec:methods}

We have solved 
the governing equations 
using a number of different codes.
The first is the time-integration code written by Patankar \cite{Patankar}
for equations \eqref{eq:goveqns}--\eqref{eq:bcs}.
The temporal discretization uses
the backwards differentiation formula for the time derivative,
the explicit Adams-Bashforth formula for the advective and buoyancy terms,
and implicit evaluation of the diffusive terms, which
leads to the following form for timestepping equation \eqref{eq:goveqns}:
\begin{eqnarray}
\frac{1}{2\dt}\left[3 
\left(\begin{array}{c}\bU \\ T \end{array}\right)^{t+\dt}
-4 \left(\begin{array}{c}\bU \\ T \end{array}\right)^t
+\left(\begin{array}{c}\bU \\ T \end{array}\right)^{t-\dt}
\right] \nonumber\\
+ 2 \left(\begin{array}{c}(\bU\cdot\grad)\bU\\\bU\cdot\grad T\end{array}\right)^{t}
-\left(\begin{array}{c}(\bU\cdot\grad)\bU\\\bU\cdot\grad T\end{array}\right)^{t-\dt}\nonumber\\
=\left(\begin{array}{c}\frac{1}{Re} \lapv \bU \\ \frac{1}{Pr Re} \lap T
\end{array}\right)^{t+\dt} 
-\left(\begin{array}{c}\grad P \\ 0 \end{array}\right)^{t+\dt} \nonumber\\
+ 2 \left(\begin{array}{c}\frac{Ra}{Pr Re^2} T\be_z\\ 0 \end{array}\right)^{t}
-\left(\begin{array}{c}\frac{Ra}{Pr Re^2} T\be_z\\ 0 \end{array}\right)^{t-\dt}
\nonumber\end{eqnarray}
The spatial discretization uses centered finite differences.
A staggered grid is used, with pressure and temperature represented 
on the nodes and the velocity components between the nodes.
A timestep $\Delta t=10^{-3}$ and a resolution in the radial and 
axial direction of $82\times 82$ were used.

Time integration has also been carried out with a second code,
which uses the streamfunction-vorticity formulation
\eqref{eq:eqstrm}--\eqref{eq:bcstrm}, a second-order 
finite difference scheme in space, and an alternating-direction-implicit 
scheme in time \cite{LMW_JFM_01}.
A $51\times 51$ grid with a typical timestep of 
$\Delta s=10^{-4}$ were found to be adequate for obtaining converged results. 

We also carried out bifurcation and linear stability analysis.
We followed branches via Newton's method and calculated eigenvalues
via ARPACK.
Several different spatial representations have been used.
One code uses the same spatial discretization -- i.e. the
streamfunction-vorticity formulation and the same spatial resolution -- 
as the time-integration code mentioned in the previous paragraph. 
The Jacobian of the matrix built in the last
step of Newton iteration is directly used in ARPACK in shift-invert mode. 
The branches are followed using an arclength continuation as described in 
\cite{Doedel} and the practical implementation of bordered matrix inversion
can be found in \cite{Govaerts}.

The second code \cite{Maria_07} with which we carried out bifurcation and
linear stability analysis uses the primitive variable formulation and a
pseudo-spectral spatial discretization.  The number of Chebyshev polynomials
(or Gauss-Lobatto collocation points) used is 15 in each of the $r$ and $z$
directions.

The quantitative results which we present have been obtained 
by two or more of the codes described above, and have also 
been verified by varying the resolution.

\section{Thresholds}
\label{sec:thresholds}
\begin{figure}[h*]
\vspace*{-1.1cm}
\centerline{
\psfig{file=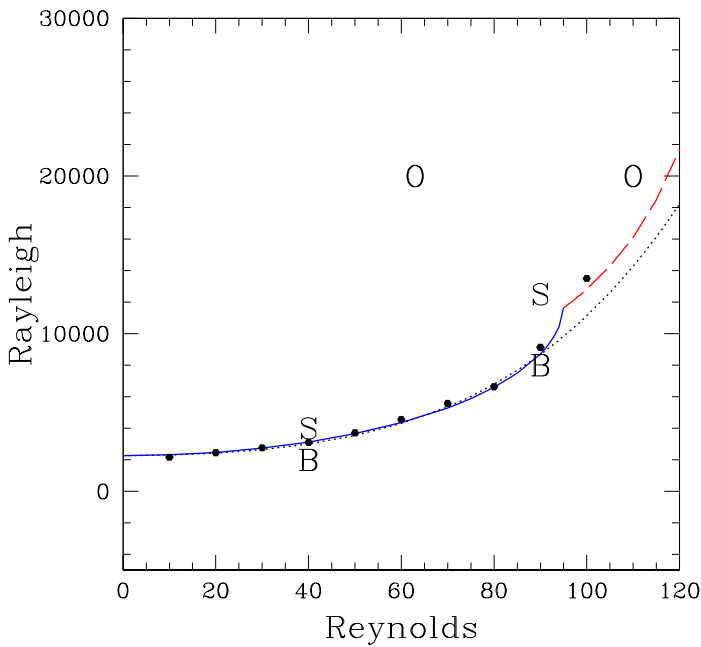,width=10cm}}
\vspace*{-0.6cm}
\caption{(Color online) Convection thresholds as a function of Reynolds number.  Solid blue
  curve: transition to steady convection.  Dashed red curve: transition to
  oscillatory convection.  Dotted black curve: fit to even fourth-order
  polynomial (\ref{eq:fit}).  Points: thresholds obtained from time-dependent
  code.  Letters indicate parameter values for flows presented in figures
  \ref{fig:conductive_Re40} and \ref{fig:conductive_Re90} (B, basic state),
  \ref{fig:convective_Re40} and \ref{fig:convective_Re90} (S, steady
  convection), and \ref{fig:re110_visu} and \ref{fig:re63_visu} (O,
  oscillatory convection).}
\label{fig:thresholds}
\end{figure}

In the absence of rotation, i.e.~$Re=0$, the governing equations 
have as a solution the conductive state, in which the
fluid is stationary and the temperature varies linearly in the vertical
direction:
\begin{equation}
\bU = 0, \qquad T = 1-z
\label{eq:conductive}\end{equation}
Although the conductive solution exists for all $Ra$, it loses stability at
some critical value $Ra_c$ to a convective solution, in which fluid motion
ensues, breaking the Boussinesq reflection symmetry \eqref{eq:reflection}.  In
the von K\'arm\'an system, as mentioned in section \ref{sec:problem}, Ekman
pumping leads to recirculating regions for any non-zero value of
$Re$. However, for a fixed value of $Re$, there is still a transition at a
well-defined $Ra_c$, at which one solution loses stability to another.  By
analogy with the non-rotating case, we will call these the {\it basic} and
{\it convective} solutions, despite the presence of fluid motion (and thus of
convective heat transfer) for $Ra< Ra_c$, and we will continue to refer to the
transition between them as the onset of convection.

We begin by showing the critical Rayleigh number for onset of axisymmetric
convection as a function of Reynolds number.  For $Re < 95$, the bifurcation
is to a steady state (i.e.~the critical eigenvalue is real), and for $Re >
95$, the bifurcation is to an oscillatory state (i.e.~the critical eigenvalue
is complex).  The thresholds in figure \ref{fig:thresholds} are obtained by
calculating the eigenvalues of the linearized problem for fixed values of
$(Re,Ra)$ and interpolating to find the value $Ra_c(Re)$ at which the
eigenvalue or its real part crosses zero. Thresholds were also obtained
independently by fitting decay rates from the time-stepping code to
exponentials and extrapolating these to zero.

Away from where the bifurcation changes its nature from steady to oscillatory,
$Ra_c$ should be a smooth function of $Re$. In fact,
$Ra_c$ must be an even function of $Re$, since reversing $Re$
means reversing the direction of rotation of the upper and
lower disks, an operation which leaves all dynamical properties
unchanged, as was argued \cite{Muller_92} in the context 
of Rayleigh-B\'enard convection with throughflow.
We fit $Ra_c$ over the range of the steady bifurcation, 
$0 \leq Re \leq 94$, to an even polynomial, obtaining:
\begin{equation}
Ra_c \approx \RacRezero  +(0.6243 \times Re)^2 + (0.0840 \times Re)^4
\label{eq:fit}
\end{equation}
We see that the von K\'arm\'an flow
stabilizes the system against thermal convection. 
%
We contrast this with results from other mixed thermal/shear systems.  In
plane Poiseuille and Couette flow, the threshold for onset of longitudinal
rolls is unaffected by the shear \cite{CB_91,CB_92}.  (In our system, the
rolls can be seen as longitudinal, since the axes of concentric rolls are
azimuthal, as is the velocity of the bounding disks.)  In Taylor-Couette flow
(measured by the Taylor number $Ta$) with a radial temperature gradient
(measured by the Grashof number $Gr$), the role of rotation and heating are
reversed from our system.  Any radial temperature gradient causes large-scale
motion (a role played by $Re$ in our system) and a bifurcation to Taylor
vortices is seen at a finite rotation rate (a role played by $Ra$ in our
system).  In experiments on this system \cite{Lepiller}, $Ta_c$ decreases with
$Gr$, in contrast to the increase in $Ra_c$ with $Re$ in our case.

The threshold value $Ra_c(Re=0)= \RacRezero  $ agrees well with those previously
reported for the onset of convection for
insulating sidewalls and aspect ratio $\Gamma=1$, notably
2260  \cite{ChaSan71,BueCat},
2300 ($\Delta Ra_c/Ra_c$=1.8\%) \cite{WanKuhRat}, 
2241 ($\Delta Ra_c/Ra_c$= 0.8\%) \cite{TouHadHen}, and
2250 ($\Delta Ra_c/Ra_c$= 0.4\%) \cite{Boronska_JFM}.
These authors all found the most unstable mode to be axisymmetric 
for cylindrical convection at this aspect ratio.

Varying $Re$ and investigating the thresholds of non-axisymmetric eigenmodes
shows that the critical azimuthal wavenumber is $m=2$ for $Re \geq
\RetranstriD$.  At $Re=0$, the primary axisymmetric convective branch becomes
unstable to an $m=2$ perturbation near $Ra=3000$ \cite{WanKuhRat,TouHadHen}.
For the pure von K\'arm\'an flow, i.e. with $Ra$ set to zero and $Re$
gradually increased from zero, the first instability occurs near $Re = 300$
to an $m=2$ mode \cite{Nore_04}.  We nonetheless restrict our consideration
here to the axisymmetric bifurcations and eigenmodes. There are a number of
reasons for doing so. First, an understanding of the axisymmetric problem is
important for completeness: branches which are unstable may nonetheless play a
role in the dynamics.  Secondly, the axisymmetric problem displays a number of
fascinating phenomena, both linear and nonlinear, as we will see.  Because of
this, it has served as a test case for a numerical method using a reduced
model \cite{Maria_09}; a presentation of the results from fully resolved
computations is a necessary complement to this study. Finally, additional
physical effects, e.g.~magnetic fields, can stabilize the axisymmetric
configuration \cite{Houchens}.

\section{Steady convective states}
\label{sec:steady}

\begin{figure}[t]
\includegraphics[width=8cm]{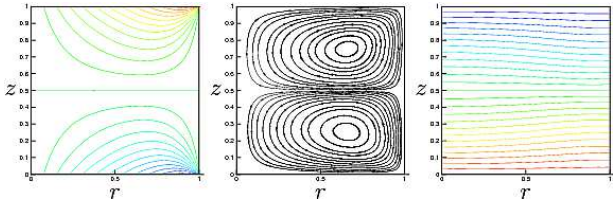}
\caption{(Color online) 
Basic flow at $Re=40$, $Ra=2000$.
From left to right: $U_\theta$, $\Psi$ and $T$.
Ranges for $U_r$ and $U_z$ are $[-0.05,0.05]$ and $[-0.04,0.04]$, respectively
}
\label{fig:conductive_Re40}
\end{figure}
\begin{figure}
\includegraphics[width=8cm]{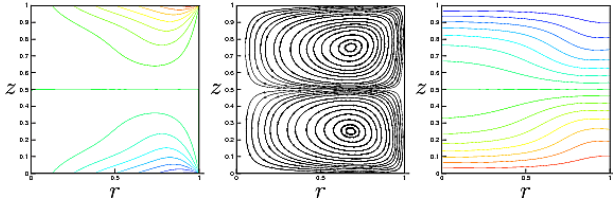}
\caption{(Color online) Basic flow at $Re=90$, $Ra=8000$.
From left to right: $U_\theta$, $\Psi$ and $T$.
Ranges for $U_r$ and $U_z$ are $[-0.11,0.11]$ and $[-0.07,0.07]$,
respectively.
}
\label{fig:conductive_Re90}
\end{figure}
\begin{figure}[t]
\includegraphics[width=8cm]{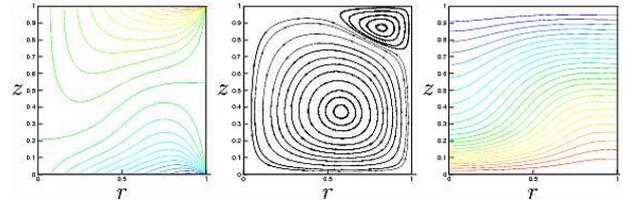}
\caption{(Color online) Convective state at $Re=40$, $Ra=4000$.
From left to right: $U_\theta$, $\Psi$ and $T$.
Ranges for $U_r$ and $U_z$ are $[-0.11,0.19]$ and $[-0.29,0.11]$, 
respectively.
}
\label{fig:convective_Re40}
\end{figure}
\begin{figure}[h*]
\includegraphics[width=8cm]{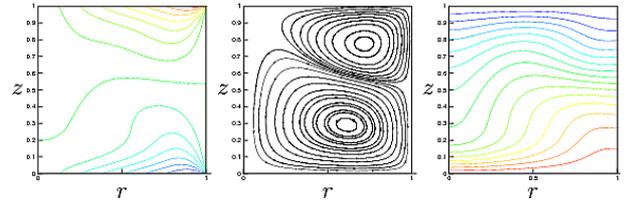}
\caption{(Color online) Convective state at $Re=90$, $Ra=12500$.
From left to right: $U_\theta$, $\Psi$ and $T$.
Ranges for $U_r$ and $U_z$ are $[-0.15,0.18]$ and $[-0.24,0.10]$, 
respectively.
}
\label{fig:convective_Re90}
\end{figure}

\begin{figure*}[htb]
\centerline{
\includegraphics[width=17cm]{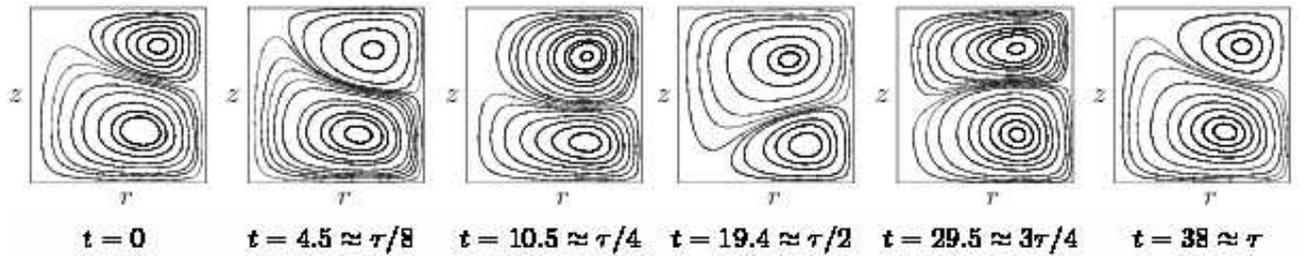}}
\caption{Near-sinusoidal limit cycle at $Ra=20\,000$, $Re=110$.  $\Psi$ is
  shown at times near $0$, $\tau/8$, $\tau/4$, $\tau/2$, $3\tau/4$, $\tau$
  indicated on timeseries in figure \ref{fig:timeseries}.  Times are given in
  units of $1/\Omega$.  The overall limit cycle has reflection symmetry: the
  states in the second half of the cycle are related by reflection symmetry to
  those in the first half of the cycle.  The vortex in the upper right corner
  is smaller during the first half of the cycle and larger during the second
  half.  The streamfunction contours are not equally spaced, but instead
  chosen to illustrate topological features of the flow.}
\label{fig:re110_visu}
\end{figure*}
\begin{figure*}[htb]
\centerline{
\includegraphics[width=17cm]{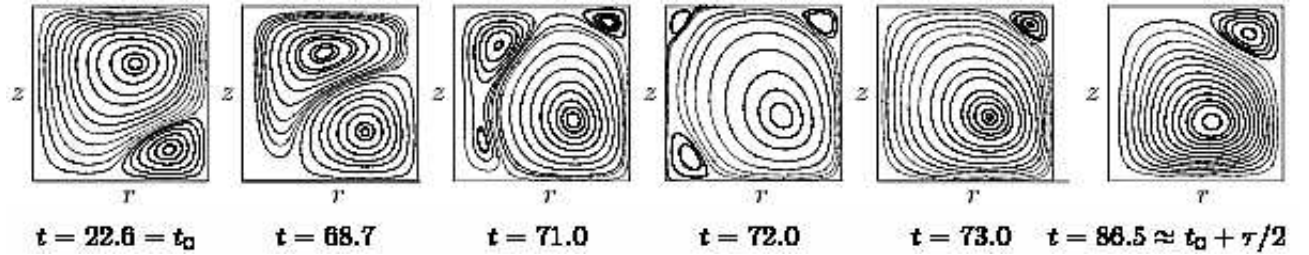}}
\caption{Near-heteroclinic limit cycle at $Ra=20\,000$, $Re=63$.
$\Psi$ is shown at times indicated on timeseries in figure \ref{fig:timeseries}:
$t=22.6$; four instants in 
$68\leq t \leq 73$, during which the flow changes a great deal;
and $t=86.5$, approximately a half-period $\tau/2\approx 57.5$ after $t=22.6$.
The vortex in the lower right corner grows and the vortex in the
upper left corner folds and then divides into three small vortices.
Two of these disappear, leaving a small vortex in the upper right corner.
As in figure \ref{fig:re110_visu}, the states in the second half of the 
cycle are related by reflection symmetry to those in the first half of 
the cycle and the streamfunction contours are chosen to illustrate 
topological features of the flow.}
\label{fig:re63_visu}
\end{figure*}

We now discuss the nature of the steady flows below and above the convection
threshold.  The basic state, which is the analogue of the conductive state in
the presence of von K\'arm\'an flow, is shown in figures
\ref{fig:conductive_Re40} and \ref{fig:conductive_Re90} for $Re=40$, $Ra=2000$
and for $Re=90$, $Ra=8000$ respectively. The azimuthal velocity increases
gradually from negative in the lower half to positive in the upper half of the
cylinder, following the counter-rotating disks. It is singular at the corners
$r=1$, $z=0,1$ since the boundary conditions are discontinuous where the
stationary cylinder meets the rotating disks.  
The large recirculating cells caused by Ekman pumping 
are prominent features in the $(r,z)$ plane,
carrying fluid radially outward along both rotating disks at $z=0,1$,
then along the bounding cylinder, inwards along the midplane $z=1/2$, 
and back towards the two disks along the axis.  
The azimuthal velocity and recirculating cells combine to yield 
fluid trajectories which are toroidal.  

We may compare the basic flows in figures \ref{fig:conductive_Re40} and
\ref{fig:conductive_Re90}.  For $Re=90$, the temperature field is noticeably
different from the linear conductive profile \eqref{eq:conductive}.  The
recirculating cells are stronger, with a maximal velocity of $|U_r^{\rm
  max}|=0.11$, as compared to $|U_r^{\rm max}|=0.04$ for $Re=40$.  These
stronger recirculating cells exert more influence on the temperature field.
At $r=1$, the isotherms are deviated towards the midplane as warm fluid 
converges upwards from the lower disk and cold fluid downwards from the upper
disk. At $r=0$, the isotherms are deviated again, here outwards towards the
upper and lower disks.  The azimuthal velocity gradients for $Re=90$ are more
concentrated near the disks.
The basic states shown in figures \ref{fig:conductive_Re40}
and \ref{fig:conductive_Re90} are reflection-symmetric about the midplane:
each is invariant under $\kappa$, as defined in \eqref{eq:reflection}.

The transition to convection breaks the reflection symmetry.
In the absence of rotation, the transition occurs at $Ra= \RacRezero$, 
as stated in section \ref{sec:thresholds}, and leads to 
a single roll which occupies the entire cavity.
For $Re=40$, transition occurs at $Ra_c=\RacReforty$ and the resulting convective 
state, shown in figure \ref{fig:convective_Re40} for $Ra=4000$, has 
one small roll (due to the weak von K\'arm\'an flow) in the upper right 
corner in addition to the main large convection roll.
%
For $Re=90$, the onset of convection occurs at $Ra=\RacReninety$ and the resulting
convective state is shown in figure \ref{fig:convective_Re90} at $Ra=12\,500$.
The influence of the counter-rotating disks is stronger than it is at $Re=40$
and so the recirculating rolls, resulting from the combined effects of
Ekman pumping (which favors two equal rolls) and convection (which favors 
a single roll), are more equal in size.  
The consequences on $U_\theta$ and $T$ of these changes in $\Psi$
can also clearly be seen in figures \ref{fig:convective_Re40} and 
\ref{fig:convective_Re90}. (These in turn affect $\Psi$ via the buoyancy and
Ekman pumping effects, though to a lesser extent.)

Although the smaller cell in figures \ref{fig:convective_Re40} and 
\ref{fig:convective_Re90} is in the upper corner of the cylinder,
the pitchfork bifurcation
can produce a small cell at either the upper or lower corner.
The breaking of reflection symmetry is also manifested
by the azimuthal velocity: the $U_\theta=0$ contour intersects the axis 
near $z=0.2$ in both figures \ref{fig:convective_Re40} and
\ref{fig:convective_Re90}, in contrast to the basic flow, 
for which this contour is a straight line across the midplane at $z=0.5$.

\section{Oscillatory convective states}
\label{sec:oscill}

\begin{figure*}[htb]
\includegraphics[width=15cm]{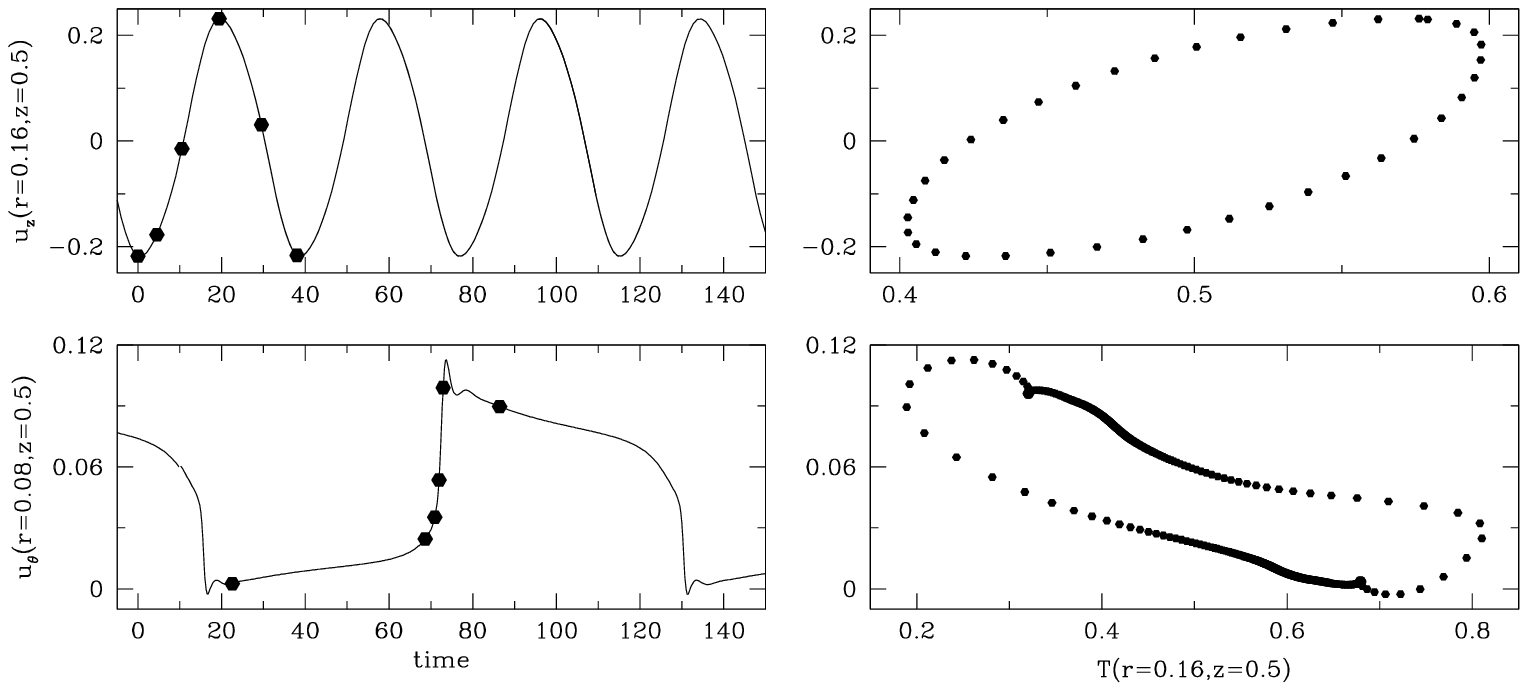}
\vspace*{-0.5cm}
\caption{Timeseries (left) and phase portraits (right) corresponding 
to limit cycles at $Ra=20\,000$. 
Dots in timeseries refer to visualizations in figures \ref{fig:re110_visu} 
and \ref{fig:re63_visu}. Dots in phase portraits are plotted
at equally spaced times so that their density reflects the rate at 
which the limit cycle is traversed.
Top: at $Re=110$, the oscillations
are regular and near-sinusoidal.
Bottom: at $Re=63$, the oscillations
have a much longer period and
consist mainly of two long plateaus with abrupt gradients between them.}
\label{fig:timeseries}
\end{figure*}
\begin{figure*}[htb]
\includegraphics[width=14cm]{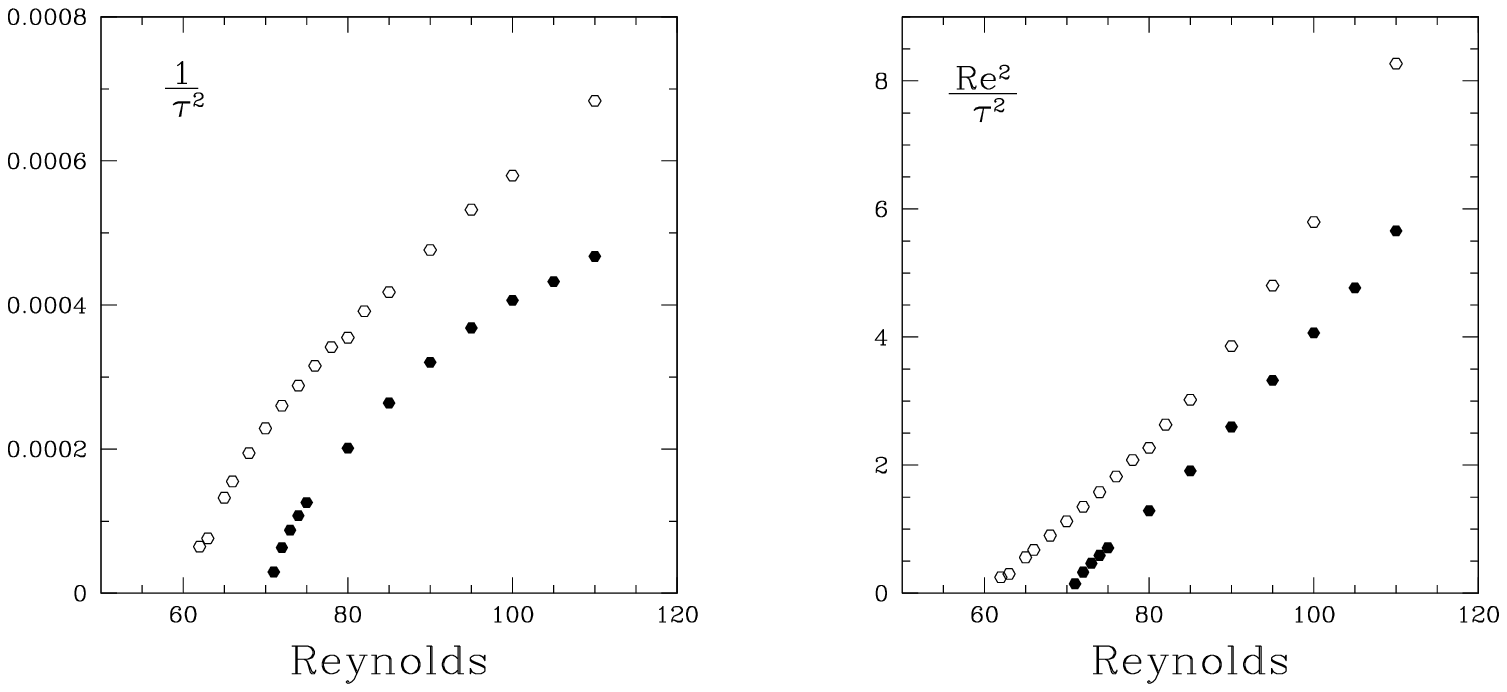}
\vspace*{-0.5cm}
\caption{Variation of square frequency with $Re$.
Left: $1/\tau^2$ for $Ra=20\,000$ (hollow dots) and for $Ra=18\,000$ (solid
dots). Right: $(Re/\tau)^2$ for $Re=20\,000$ (hollow dots) and 
for $Ra=18\,000$ (solid dots).}
\label{fig:periods}
\end{figure*}

For $Re > 95$, the transition from the basic state is a Hopf bifurcation,
creating oscillatory flows. 
We fix $Ra=20\,000$ and present two oscillatory states, or limit
cycles, one at $Re=110$ in figure \ref{fig:re110_visu} and one at
$Re=63$ in figure \ref{fig:re63_visu}. (Although for $Re=63$, the first
bifurcation as $Ra$ is increased is to steady convection, there is another
bifurcation at higher $Ra$ to oscillatory convection and the resulting limit
cycles are stable and connected to those created by the Hopf bifurcation for
$Re > 95$, as we will see in section \ref{sec:bifs}.)  We call the limit
cycles at $Re=110$ and $Re=63$ near-sinusoidal and near-heteroclinic,
respectively, which we will also explain in section \ref{sec:bifs}.

Figures \ref{fig:re110_visu} and \ref{fig:re63_visu} show 
meridional vortices jostling with one another. At Re=110
(figure \ref{fig:re110_visu}), the vortices are somewhat similar in 
size, with the vortex in the upper or lower outer corner 
alternatively becoming smaller and larger.
At Re=63 (figure \ref{fig:re63_visu}), the principle vortex remains
much larger throughout the limit cycle, with much smaller vortices
on the periphery. At $t=71.0$, there appear to be two vortices
which both extend over the entire height, suggesting an interpretation 
of this limit cycle as a competition between one and two concentric 
radial vortices. This interpretation will be discussed further in section 
\ref{sec:linear}.

Section \ref{sec:steady} showed that the transition to 
steady convection breaks the reflection symmetry, creating 
two asymmetric states related by the reflection operator $\kappa$.
The transition to oscillatory convection has an analogous property:
each of the instantaneous states shown in figures \ref{fig:re110_visu} 
and \ref{fig:re63_visu} is asymmetric, and the second half of 
each limit cycle is related to the first half by reflection, i.e.:
\begin{equation}
(U_r,U_\theta,U_z,T)(t+\tau/2) = \kappa(U_r,U_\theta,U_z,T)(t)
\end{equation}
where $\tau$ is the oscillation period.

To better understand the limit cycles in figures \ref{fig:re110_visu}
and \ref{fig:re63_visu}, we plot various scalar quantities -- $U_z(0.16,0.5)$,
$U_\theta(0.08,0.5)$ and $T(0.16,0.5)$ -- as a function of time and of each
other to create the timeseries and phase portraits of figure
\ref{fig:timeseries}.  Figure \ref{fig:timeseries} clearly shows the very 
different character of the limit cycles
at $Re=110$ and $Re=63$.  At $Re=110$, the
timeseries is fairly close to harmonic, and the phase portrait shows that the limit
cycle is traversed at a fairly constant speed.  In contrast, at $Re=63$, the
timeseries shows two phases during which change is very slow, punctuated by
abrupt transitions.  The phase portrait corroborates this: two regions of the
limit cycle show a dense accumulation of points, indicating a slow traversal
of these regions.

We have seen that the period at $Re=63$ is about three times
that at $Re=110$; it in fact approaches infinity as $Re$ is decreased.
(The approach by $\tau$ to infinity implies that it is extremely sensitive to
any change in the physical or numerical parameters. Thus, values of $\tau$
are subject to a great deal of uncertainty.)
This is demonstrated in figure \ref{fig:periods}, which
plots the dependence of the square frequency (inverse square period) on $Re$
for $Ra=20\,000$ and for $Ra=18\,000$,
for two different scalings: with the disk rotation period $1/\Omega$ (left)
and with the viscous diffusion time $R^2/\nu$ (right).

\section{Bifurcation Diagram}
\label{sec:bifs}

\begin{figure}[htb]
\centerline{
\includegraphics[width=9cm]{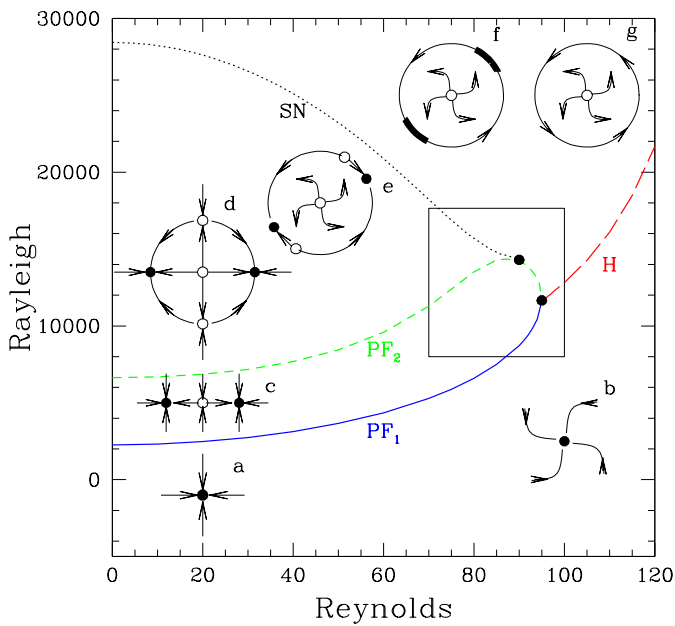}}
\vspace*{-0.5cm}
\caption{(Color online) Curves of bifurcation points in the $(Re,Ra)$ plane
showing first pitchfork (blue, solid, $PF_1$), 
second pitchfork (green, short-dashed, $PF_2$), 
Hopf (red, long-dashed, $H$), 
and saddle-node (black, dotted, $SN$) bifurcations.
Dots indicate codimension-two points,
where curve $PF_2$ meets curves $SN$ (90,14\,300) and $H$ (95,\RaTakens).
An enlargement of the region inside the square is shown in figure 
\ref{fig:codim2_zoom}.
Phase portraits shown as insets.
a, b) Below $PF_1$ and $H$, the only solution is the stable basic flow, 
which is a node (a) or a spiral focus (b).
c) Between $PF_1$ and $PF_2$, the basic flow is 
unstable and there exist two stable asymmetric convective states.
d, e) Between $PF_2$ and $SN$, there exist five states:
the unstable basic flow and two stable and two unstable convective states.
f, g) Between $SN$ and $H$, the unstable basic flow is surrounded by 
a stable limit cycle, which is near-heteroclinic near $SN$ (f)
and near-sinusoidal near $H$ (g).}
\label{fig:codim2}
\end{figure}

\begin{figure*}[htb]
\centerline{
\includegraphics[width=18cm]{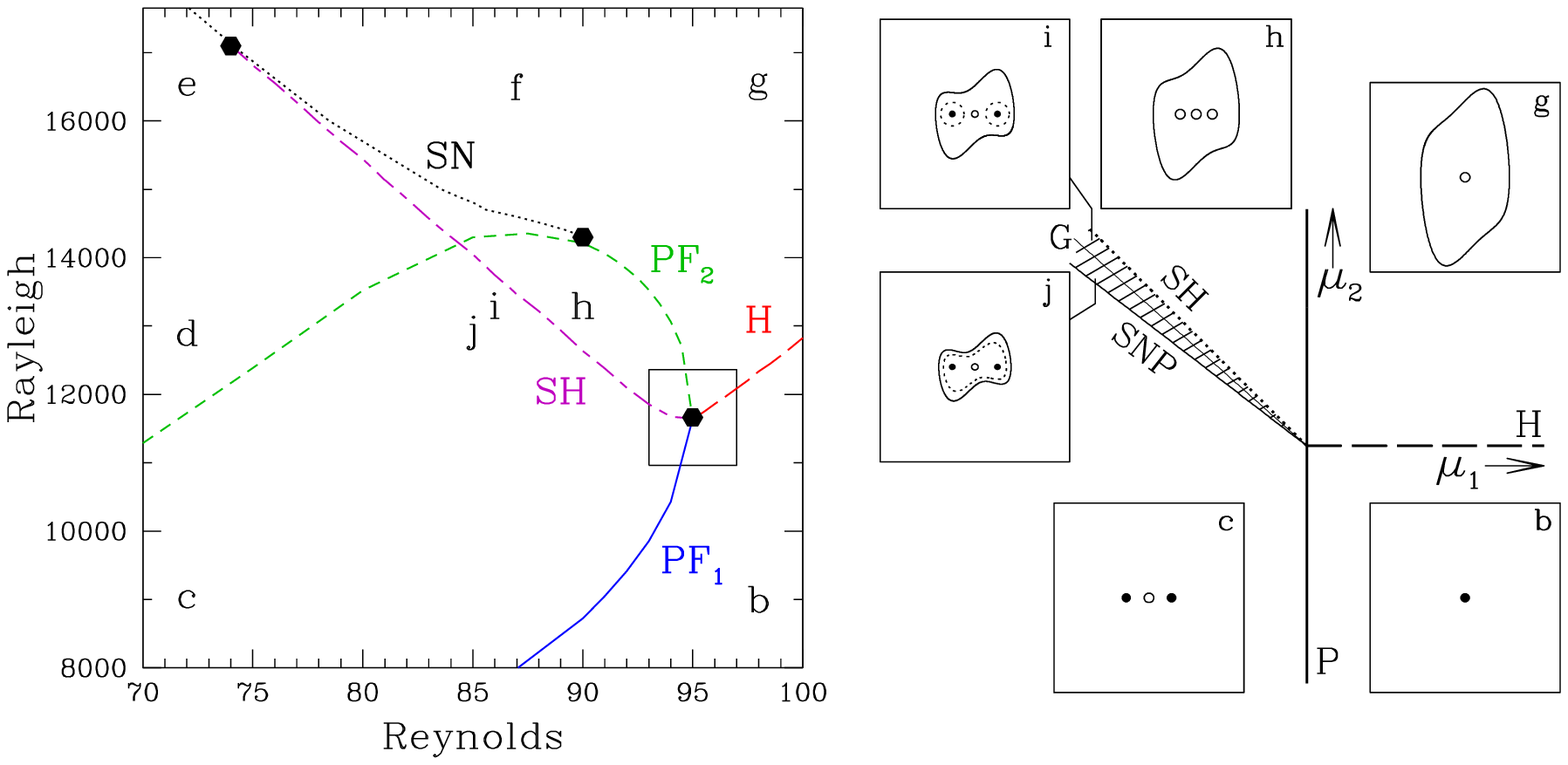}}
\caption{(Color online) Left: enlargement of square in figure \ref{fig:codim2}.  
The additional curve (violet, dash-dotted) indicates the secondary 
Hopf bifurcation $SH$ which exists between codimension-two points 
$(95,\RaTakens)$ (where it meets $H$) and $(74,17\,100)$ (where it meets $SN$).
Labels correspond to those of figure \ref{fig:codim2}.  Behavior in the square
is described by the TB normal form \eqref{eq:TB}.  \\ Right: behavior of the
TB normal form \eqref{eq:TB}. Shown are the pitchfork bifurcation $P$, Hopf
bifurcation $H$, secondary Hopf bifurcation $SH$, as well as the saddle-node
of periodic orbits $SNP$ and the gluing bifurcation $G$ which are contained in
the unfolding of the TB point.  Phase portraits b),
c), g) as in figure \ref{fig:codim2}.  h) Between $P$ and $SH$ a stable limit
cycle surrounds the unstable basic and convective flows. i) Between $SH$ and
$G$, small unstable limit cycles surround each stable convective
state.  j) Between $G$ and $SNP$, two large limit cycles, one stable and one
unstable, surround the three steady states. Hatched region indicates 
bistability between limit cycle and steady states.}
\label{fig:codim2_zoom}
\end{figure*}

We have mentioned several features of the oscillatory states in 
section \ref{sec:oscill}: \\
$\bullet$ The small vortex does not remain in the upper or lower corner,
as it does for the steady convective states, but alternates between the two
locations, as shown in figures \ref{fig:re110_visu} 
and \ref{fig:re63_visu}.\\
$\bullet$ The character of the oscillations varies substantially, 
from near-sinusoidal to relaxational (plateaus punctuated by fast changes)
as shown in figure \ref{fig:timeseries}.\\
$\bullet$ The period varies substantially with changes in $Re$ or $Ra$, 
as shown in figure \ref{fig:periods}.\\
All of these features are typical of the bifurcation scenario 
we will now describe, and which is illustrated in figure \ref{fig:codim2}.

For $Re<95$, the threshold for convection, already shown in figure
\ref{fig:thresholds}, corresponds to a pitchfork bifurcation $PF_1$.  For $Ra$
below $PF_1$, i.e.~inset (a) in figure \ref{fig:codim2}, the only solution is
the basic flow, an example of which is given in figure
\ref{fig:conductive_Re40}.  Proceeding clockwise, at $PF_1$, the basic state
loses stability and gives rise to two symmetrically-related convective states,
as illustrated in inset (c) and figure \ref{fig:convective_Re40}.  A second
pitchfork bifurcation, $PF_2$, occurs at a higher value of $Ra$.  For $Ra$
between curves $PF_2$ and $SN$, in regions (d) and (e), there exist two
additional steady states, both unstable and also related by the symmetry
operation \eqref{eq:reflection}.  The trajectories leaving the two unstable
states terminate on the two stable states, forming an invariant set that
resembles a circle or ellipse.  Leaving $PF_2$ and approaching $SN$, as shown
in (e), the two stable and unstable states approach one other along this
invariant set, finally merging and annihilating one another in two 
symmetrically related saddle-node bifurcations at $SN$.

Exactly at $SN$, trajectories spend an infinite period of time at the two
locations at which the stable and unstable states merged, forming a limit
cycle whose period is infinite: a heteroclinic cycle.  For $Ra$ above $SN$,
but sufficiently nearby, as in inset (f), much of the time during the limit
cycle is spent in the neighborhood of the former fixed points, as shown for
$Re=63$, $Ra=20\,000$ in figure \ref{fig:re63_visu} and in the bottom half of
figure \ref{fig:timeseries}.  Leaving $SN$, by increasing $Ra$ or $Re$, as in
inset (g), the time spent in the vicinity of these points shortens, and the
period decreases, as shown for $Re=110$, $Ra=20\,000$ in figure
\ref{fig:re110_visu} and in the upper half of figure \ref{fig:timeseries}.
Finally, for $Re>95$, curve $H$ corresponds to a Hopf bifurcation, below
which the basic state is a stable spiral focus, as in inset (b).

Although the saddle-node bifurcation signals the merging of two pairs of
steady states, it results in a limit cycle because of the connections between
these steady states. (These connections in turn result from the formation of
the steady states via two successive pitchfork bifurcations.)  Under these
circumstances, the saddle-node bifurcation is called a SNIPER, for Saddle-Node
In a PERiodic orbit, or Saddle-Node Infinite PERiod; other names are a
saddle-node homoclinic, a SNIC (Saddle-Node on Invariant Circle), or an
Andronov bifurcation \cite{Andronov,Kuznetsov}.  It can be shown that, near a
SNIPER bifurcation, the period $\tau$ of the limit cycle varies like
\begin{equation}
\frac{1}{\tau^2} \sim \mu-\mu_c
\label{eq:lindep}\end{equation}
where $\mu-\mu_c$ can be either $Ra-Ra_{\rm SN}$ for fixed $Re$ 
or $Re-Re_{\rm SN}$ for fixed $Ra$ or any combination of the two.
Although the square frequency
is proportional to the distance from threshold sufficiently near the
threshold for any scaling, figure \ref{fig:periods} shows that, 
for our case, the linear dependence \eqref{eq:lindep}
holds over a wider range of $Re$ when $\tau$ is scaled by 
the viscous diffusion time (right) than when it is scaled 
by the disk rotation period (left).

Figure \ref{fig:codim2} shows two codimension-two points. At (95,\RaTakens),
curves $H$, $PF_1$ and $PF_2$ meet in a Takens-Bogdanov (TB) point.  At
(90,14\,300), curve $SN$ terminates on $PF_2$ at a hysteresis point, where the
pitchfork bifurcation changes from supercritical to subcritical.
(Because the TB point was calculating using eigenvalue computations rather
than by specialized algorithms such as those in \cite{Govaerts}, we have been
able to determine its location only to within about 0.5\% in $Re$ and 5\% in
$Ra$.)
Between these two points, figure \ref{fig:codim2} shows
curve $PF_2$ separating region (c), in which the unstable basic state
coexists with two stable steady states, and region (f,g) in which it coexists
with a stable limit cycle. The pitchfork bifurcation $PF_2$ cannot bring about
such a transition.  (The other paths between (c) and (f,g), either clockwise
through $PF_2$, region d/e and $SN$; or counter-clockwise through $PF_1$,
region a/b and $H$, can bring about this transition.)  This means that figure
\ref{fig:codim2} is incomplete.  The solution is to be found in the normal
form of the Takens-Bogdanov codimension-two point in the presence of 
reflection symmetry \cite{Arnold,Kuznetsov}:
\begin{subequations}
\begin{eqnarray}
\frac{dx}{dt} &=& y \\
\frac{dy}{dt} &=& -\mu_1 x + \mu_2 y - \delta x^3 -x^2 y
\end{eqnarray}
\label{eq:TB}
\end{subequations}
where $\delta=\pm 1$. We take $\delta=+1$, since it is this 
choice which reproduces the phenomena we see in the fluid-dynamical simulation.

\begin{table}[htb]
\begin{tabular}{|c|c|l|}
\hline
Steady state & Jacobian & Eigenvalues \\
\hline
$(0,0) $&$ 
\left(\begin{array}{cc} 0 & 1 \\ 
-\mu_1 & \mu_2
\end{array}\right) $&$
\lambda_\pm = \frac{1}{2}\mu_2\pm
\sqrt{\frac{1}{4}\mu_2^2-\mu_1}$\\
$(\pm\sqrt{-\mu_1},0) $&$ 
\left(\begin{array}{cc} 0 & 1 \\ 
2\mu_1 & \mu_2\mu_1
\end{array}\right) $&$
\lambda_\pm = \frac{1}{2}(\mu_2+\mu_1)$\\
&&$\quad
\pm\sqrt{\frac{1}{4}(\mu_2+\mu_1)^2+2\mu_1}$\\
\hline
\end{tabular}
\caption{Properties of the Takens-Bogdanov normal form.}
\label{tab:TBtable}
\end{table}

The behavior of the normal form \eqref{eq:TB} is illustrated in 
figure \ref{fig:codim2_zoom} (right) and in table \ref{tab:TBtable}. The
$TB$ codimension-two point is located at $(\mu_1,\mu_2)=(0,0)$ and the basic
state at $(x,y)=(0,0)$.  The basic state undergoes a pitchfork bifurcation
($\lambda_+=0$), analogous to our $PF_1$ and $PF_2$, at $\mu_1=0$ and a Hopf
bifurcation ($\mathcal{R}e(\lambda_\pm)=0$), analogous to our $H$ at $\mu_2=0$
for $\mu_1>0$.  Additionally, table \ref{tab:TBtable} shows that the solutions
$(\pm\sqrt{-\mu_1},0)$ undergo a Hopf bifurcation at $\mu_2=-\mu_1$ for
$\mu_1<0$.  Indeed, we have calculated a secondary Hopf bifurcation $SH$ for
the hydrodynamic system, and indicate it in figure \ref{fig:codim2_zoom}.
This bifurcation is subcritical and the limit cycles it generates are
unstable.  Arnold \cite{Arnold,Kuznetsov} has shown that system \eqref{eq:TB}
contains two additional bifurcation curves emanating from the codimension-two
point $(\mu_1,\mu_2)=(0,0)$, one a gluing bifurcation, $G$, and the other a
saddle-node of periodic orbits, $SNP$.  Although the curves $SNP$ and $G$
correspond to global bifurcations and cannot be calculated by examination of
the eigenvalues in table \ref{tab:TBtable}, they have been determined
\cite{Kuznetsov} to be $\mu_2=-0.752\,\mu_1$ for $SNP$ and $\mu_2=-0.8\,\mu_1$
for $G$.  (We have computed neither for our hydrodynamic system.)  Together,
bifurcations $SNP$, $G$, $SH$ and $P$ accomplish the transformation in the
small square of figure \ref{fig:codim2}, between the three steady states of
region (c) to the stable limit cycle of region (g).

We describe the bifurcations and phase portraits in figure
\ref{fig:codim2_zoom}, proceeding around the TB point in the counter-clockwise
direction.  In region (b), the only solution is the stable basic state.  The
Hopf bifurcation $H$ destabilizes the basic state and generates a stable limit
cycle (g).  The pitchfork bifurcation $P$ is subcritical, and generates two
unstable asymmetric states while reducing the number of unstable directions of
the basic state from two to one (h).  The secondary Hopf bifurcation $SH$ is
also subcritical, stabilizing the two asymmetric states while creating
unstable limit cycles encircling each one (i).  The gluing bifurcation $G$
joins these two limit cycles to the symmetric state in a figure-eight which,
afterwards, form a single unstable limit cycle encircling all three steady
states (j).  Curves $SH$ and $SNP$ delineate a narrow wedge in which there is
bistability between the limit cycle and the steady convective states.  The
stable and unstable limit cycles annihilate in a saddle-node of periodic orbits
$SNP$, leaving only the unstable basic flow and the stable asymmetric
steady states (c).  In crossing the pitchfork bifurcation $P$, we return to
the single stable symmetric state of b).
Figure \ref{fig:codim2_zoom} shows a third codimension-two point
where $SH$ and $SN$ meet at (74,17\,100). Other codimension-two points
which we have not located must mark the ends of curves $G$ and $SNP$ as well.

Although the analysis surrounding the TB point is important for completeness,
we emphasize that, over most of the $(Re,Ra)$ plane, it is the 
SNIPER bifurcation which marks the boundary between 
steady and oscillatory convection; other bifurcations come into play 
only between (95,\RaTakens) and (74,17\,100).

\begin{figure*}
\centerline{
\includegraphics[width=15cm]{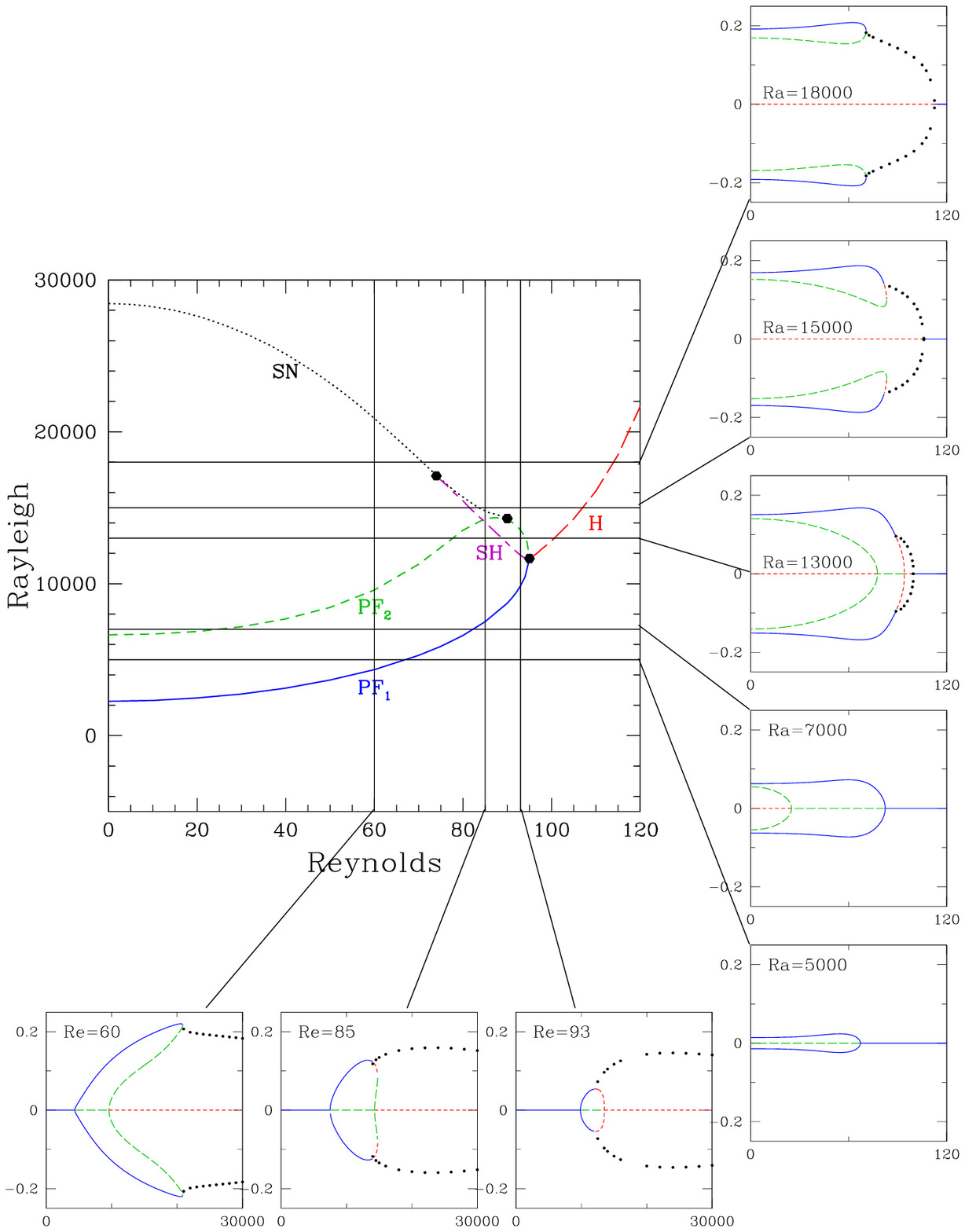}}
\caption{(Color online) Bifurcation diagrams along lines in central $(Re,Ra)$ diagram.
  $T(1/2,1/2)$ is plotted as a function of $Ra$ for fixed $Re$ in the diagrams
  along the bottom, and as a function of $Re$ for fixed $Ra$ in the diagrams
  in the column on the right.  Diagrams contain steady branches with zero 
  (blue, solid), one (green, long-dashed) and two (red, short-dashed) unstable
  directions.  Black dots indicate limit cycles.}
\label{fig:lines}
\end{figure*}

Figure \ref{fig:lines} shows representative bifurcation diagrams.  Branches
are labelled by their stability index, i.e. the number of eigenvalues with
positive real parts.  The stability index changes by one at a pitchfork
bifurcation (for the basic state) or a saddle-node bifurcation and by two for
a Hopf bifurcation.  We plot the temperature at a fixed point $\bar{T}\equiv
T(r=1/2,z=1/2)$ as a function of $Ra$ for fixed $Re$ along the bottom of the
figure, and as a function of $Re$ for fixed $Ra$ along the right.  To
represent limit cycles, we plot $[\frac{1}{\tau}\int_0^\tau \:\bar{T}(t)^2
  \,dt ]^{1/2}$, the $L_2$ norm of $\bar{T}$ over an oscillation period
$\tau$.  The three codimension-two points at (74, 17\,100), (90, 14\,300) and
(95,\RaTakens) and the limits of the pitchfork and saddle-node curves ($Ra=\RacRezero$
and $ \RacRezeroSN $) delimit the ranges over which different types of bifurcation
diagrams occur.\\ $\bullet$ $Re=60$ represents the large typical range $0\leq
Re \leq 74$: two successive supercritical pitchfork bifurcations generate two
pairs of steady states which are annihilated by a saddle-node (SNIPER)
bifurcation, leading to a limit cycle.  \\ $\bullet$ $Re=85$ represents
$[74,\,90]$: secondary Hopf bifurcations have been added to the
diagrams.\\ $\bullet$ $Re=93$ represents $[90,\,95]$: the second pitchfork
bifurcation has become subcritical.\\ $\bullet$ For $Re>95$ (not shown), there
are no non-trivial steady states and a Hopf bifurcation leads directly to the
limit cycle.\\ $\bullet$ $Ra=5000$ represents the range $\RacRezero \leq Ra \leq
\RacRezeroPFtwo$: a single pitchfork bifurcation leads to a pair of steady states as $Re$
decreases.\\ $\bullet$ $Ra=7000$ represents $[\RacRezeroPFtwo, \,\RaTakens]$: an additional
pitchfork bifurcation and pair of steady branches can be seen.\\ $\bullet$
$Ra=13\,000$ represents $[\RaTakens,\,14\, 300]$: a primary and two secondary
Hopf bifurcations have been added to the scenario.\\ $\bullet$ $Ra=15\,000$
represents $[14\,300,\,17\,100]$: the two pairs of non-trivial branches are no
longer connected to the trivial branch via pitchfork bifurcations, but instead
arise via saddle-node bifurcations.\\ $\bullet$ $Ra=18\,000$ represents
$[17\,100,\,\RacRezeroSN]$: the secondary Hopf bifurcations have
disappeared.\\ $\bullet$ For $Ra >  \RacRezeroSN $ (not shown), there are no
non-trivial steady states, but only a limit cycle generated by the Hopf
bifurcation.\\ (The other bifurcations depicted in the enlargement of figure
\ref{fig:codim2_zoom}, which are present for the ranges $74 \leq Re \leq 95$
and $\RaTakens \leq Ra \leq 17\,100$, are omitted from figure \ref{fig:lines}
for clarity.)

Scenarios similar to those reported in this section have been computed in
other hydrodynamic configurations.
We review these studies and list the similarities and differences between 
these observations and the current results.
In a cylinder with aspect ratio $\Gamma=5$, $Pr=10$ and thermally conducting
sidewalls \cite{Tuck_Bark_88,Tuck_Bark_89}, 
steady states are produced by two pitchfork bifurcations.
When $Ra$ is increased, a SNIPER bifurcation leads to a
pattern of five concentric rolls traveling radially inwards.  By reducing the
sidewall conductivity, other saddle-node bifurcations on the same invariant
circle appeared, creating stable steady four-roll states.  No Hopf
bifurcations or Takens-Bogdanov points were found.  Such a bifurcation may
also have been responsible for long-period oscillations observed in experimental
timeseries, but for which the flow could not be visualized \cite{Behringer}.
A SNIPER bifurcation is also found for $\Gamma=4$, $Pr=7$ in a study of
rotating convection \cite{LRM_06}. When rotation is turned on, the reflection
symmetry \eqref{eq:reflection} is broken and the pitchforks become imperfect,
but the SNIPER bifurcation survives, with a limit cycle created by one
saddle-node bifurcation instead of two simultaneous ones.

Siggers \cite{Siggers} carried out an extensive survey over $4\leq \Gamma \leq
10$, $Pr=0.1$ with (less realistic but more tractable) stress-free boundary
conditions. Like \cite{Maria_09}, the numerical method used an expansion in
eigenfunctions of the linearized problem. In \cite{Siggers}, 20--30
eigenfunctions were retained, a number sufficiently small to allow automatic
tracking of bifurcations. A phase diagram very similar in its large-scale
features to that described in this section was found: limit cycles compete
with steady states that are created by two pitchfork bifurcations and
destroyed by saddle-node bifurcations.  However, unlike in our case, the basic
state never undergoes a Hopf bifurcation like our $H$; the only Hopf
bifurcations are those, like our $SH$, which create small limit cycles, like
those in inset (i) of figure \ref{fig:codim2_zoom}, sometimes called
vaccillation.  The large limit cycle is created by neither a SNIPER nor a Hopf
bifurcation, but rather by a saddle-node of periodic orbits (as in inset (j)
of our figure \ref{fig:codim2_zoom}) or by a heteroclinic bifurcation.

SNIPER bifurcations are also observed in situations with more complicated time
dependence, in which the saddle-node bifurcations annihilate pairs of limit
cycles to initiate flow along a torus.  This is the case in a study of
Rayleigh-B\'enard convection with modulated rotation \cite{RLM_08}. The basic
state is a large-scale axisymmetric and reflection-symmetric time-periodic
flow. As $Ra$ or the modulation amplitude of the rotation is increased,
pitchfork bifurcations produce pulsating target patterns which give way to
radially travelling target patterns via a SNIPER bifurcation.
Abshagen et al. \cite{Abshagen1,Abshagen2} have carried 
out experimental and computational studies of secondary 
bifurcations in Taylor-Couette flow in a small-aspect-ratio cylinder. 
The transitions they consider are between nonaxisymmetric rotating and 
modulated rotating waves. 
However, if one of the time scales is filtered out by taking a Poincar\'e map 
and the reflection symmetry is redefined to include rotation by $\pi$ around the
cylinder axis, their results can be seen as entirely analogous to ours.
Varying the aspect ratio and the Reynolds number, the region in which 
modulated rotating waves exist is bounded by curves of SNIPER and Hopf 
bifurcations, except over a small region in which homoclinic bifurcations 
and small-amplitude modulations mediate the transition.

\section{Eigenvalues and eigenvectors}
\label{sec:linear}

\begin{figure*}
\centerline{
\includegraphics[width=16cm]{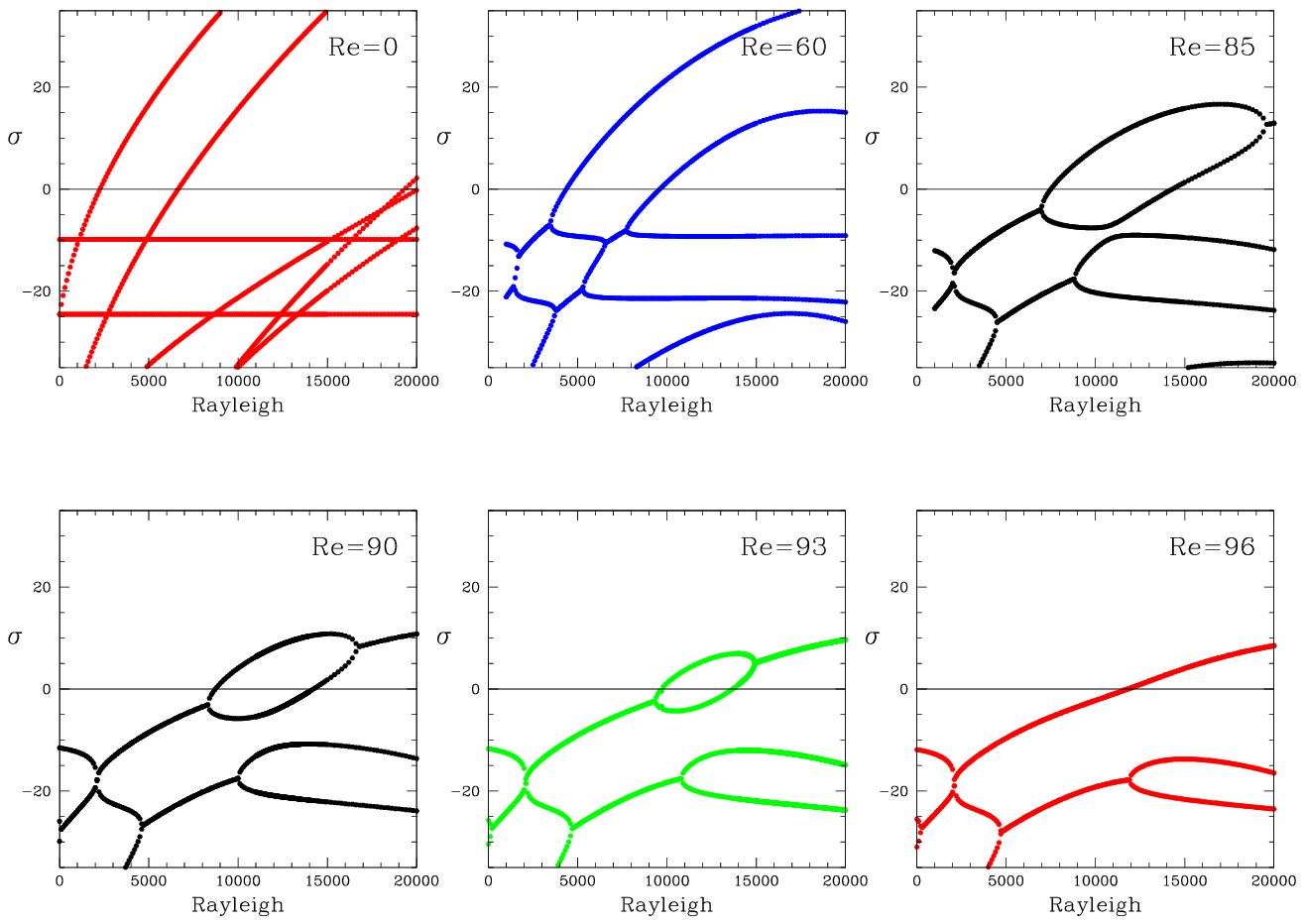}}
\caption{(Color online) Real part $\sigma$ 
of leading eigenvalues as a function of $Ra$ for six values of $Re$.
For $Re=0$, the eigenvalues are real and cross transversely.
Zero crossings at $Ra=\RacRezero$ and $Ra=\RacRezeroPFtwo$ 
correspond to pitchfork bifurcations. 
Purely thermal and azimuthal eigenvalues at $\sigma=-9.87$ and $\sigma=-24.6$, 
respectively, are independent of $Ra$.\newline
When $Re>0$, the transverse crossings become complex conjugate pairs, 
over $Ra$ intervals which widen with increasing $Re$.
By $Re=96$, the bifurcating eigenvalue is a complex conjugate pair,
leading to a Hopf bifurcation at $Ra=11\,856$.}
\label{fig:arrange_eigs}
\end{figure*}   

\begin{figure*}
\centerline{
\includegraphics[width=17cm]{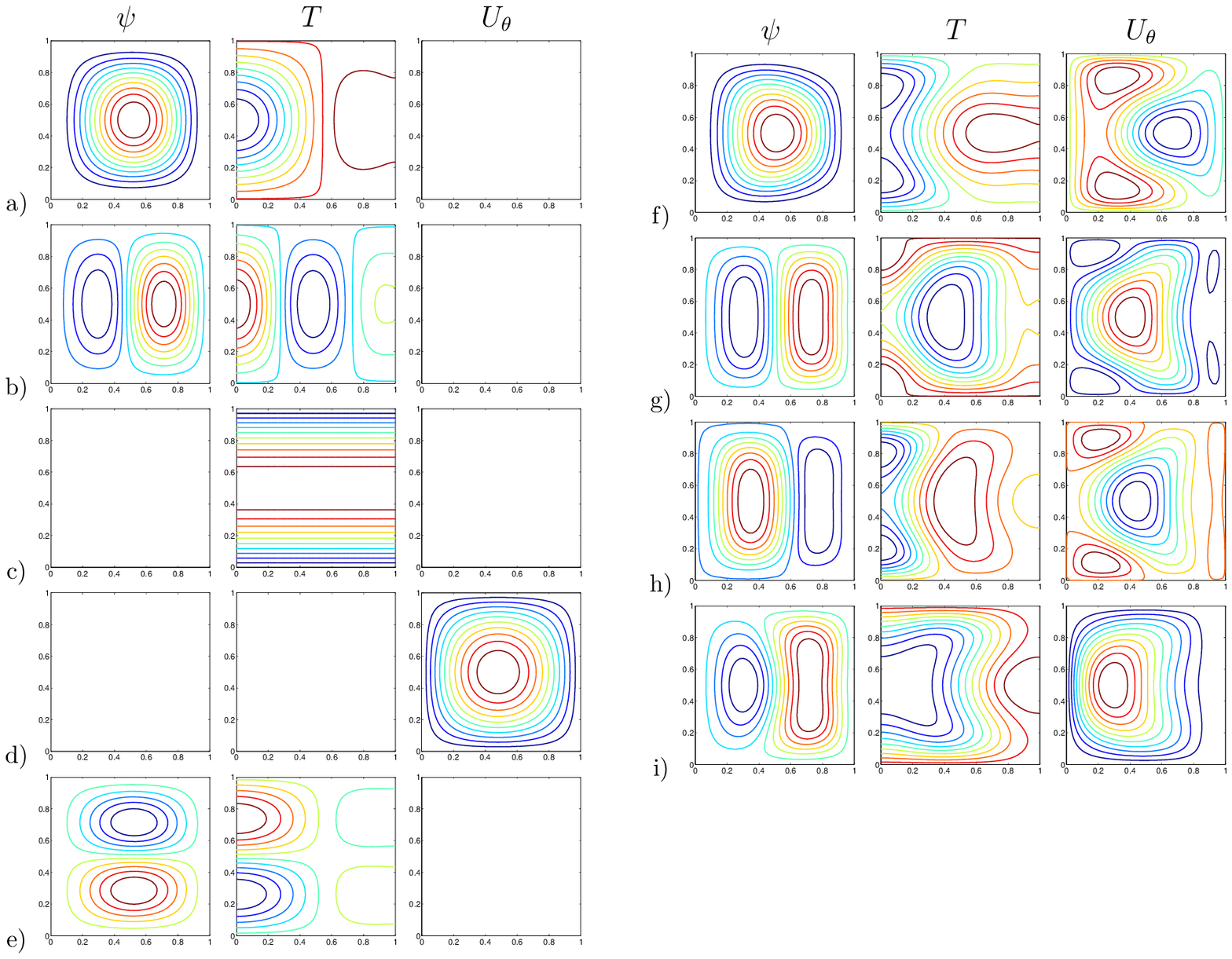}}
\caption{(Color online) Leading eigenvectors at $Ra=10\,000$. Left column: $Re=0$.
a,b) eigenvectors responsible for transition to convection with one and 
two toroidal rolls, respectively, with $\sigma=38.5$ and $\sigma=15.9$. 
c,d) thermal and azimuthal velocity eigenvectors, 
respectively, with $\sigma=-9.87\approx-\pi^2$ and 
$\sigma=-24.6 \approx -\pi^2-j_{11}^2$.
e) eigenvector with two vertically stacked rolls, with $\sigma=-21.2$.
Right column: $Re=96$.
f,g) first complex conjugate pair, with $\sigma \pm i\,\omega=-2.24\pm i \,3.45$.
h,i) second complex conjugate pair, with $\sigma\pm i\,\omega= -19.2 \pm i \,4.35$.}\label{fig:evecs}
\end{figure*}

We now focus on the behavior of the eigenvalues and eigenvectors 
of the basic flow as the Rayleigh and Reynolds numbers are varied.
Although the linear stability analysis of a problem is usually presented
before its bifurcation scenario, we will focus here on some features of 
the eigenvalues and eigenvectors that are independent of the 
bifurcation scenario.

To obtain the linear stability equations, we substitute
\begin{equation}
(\Psi,V_\theta,T)(r,z)+e^{\sigma t}(\psi,v_\theta,\Theta)(r,z)
\end{equation}
into the governing equations \eqref{eq:eqstrm}-\eqref{eq:bcstrm},  
where $(\Psi,V_\theta,T)$ is a steady solution of 
\eqref{eq:eqstrm}-\eqref{eq:bcstrm},  
and $(\psi,v_\theta,\Theta)$ is a perturbation with growth rate $\sigma$.
We note that the temperature perturbation 
$\Theta$ is unrelated to the azimuthal angle $\theta$
while the meridional velocity perturbation is 
$v_r \be_r + v_z \be_z =\be_\theta/r \times \grad\psi$.
Retaining only the linear terms leads to:
\begin{subequations}
\label{eq:eqstrmlin}
\begin{eqnarray} 
\sigma D^2\psi &=&
-\left(v_r\pd_- + v_z\pd_z\right)\:D^2\Psi \nonumber\\
&&-\left(V_r\pd_-+V_z\pd_z \right)\:D^2\psi
+\left(\lap-\frac{1}{r^2}\right)D^2\psi\nonumber\\
&&-\frac{Ra}{ Pr} \partial_r \Theta 
+2 \partial_z \left(\frac{V_\theta v_\theta}{r}\right)\\
\sigma \Theta &=&  
-(v_r\pd_r + v_z\pd_z)\: T \nonumber\\
&&- (V_r\pd_r+V_z\pd_z)\: \Theta + \frac{1}{Pr} \lap \Theta \\
\sigma v_\theta  &=&
-\left(v_r\pd_+ + v_z\pd_z\right)\:V_\theta
-\left(V_r\pd_++V_z\pd_z \right)\:v_\theta \nonumber\\
&&+ \left(\lap - \frac{1}{r^2}\right) v_\theta 
\end{eqnarray}
\end{subequations}
with homogeneous boundary conditions:
\begin{subequations}
\begin{eqnarray}
\psi = \partial_z \psi = 0 \Longleftrightarrow v_r =  v_z = 0,\nonumber\\
v_\theta = 0, \;\Theta = 0 \;\mbox{ at } z=1 \\
\psi = \partial_z \psi = 0 \Longleftrightarrow v_r =  v_z = 0,\nonumber\\
v_\theta = 0, \;\Theta = 0 \;\mbox{ at } z=0 \\
\psi = \partial_r \psi = 0 \Longleftrightarrow v_r = v_z = 0,\nonumber\\
v_\theta= 0, \;\pd_r \Theta = 0 \;\mbox{ at } r=1 \\
\psi = D^2\psi = 0 \Longleftrightarrow v_r = \pd_r v_z = 0,\nonumber\\
v_\theta= 0, \;\pd_r \Theta = 0 \;\mbox{ at } r=0
\end{eqnarray}
\label{eq:bcslin}
\end{subequations}
%
%
The Reynolds number appears via the inhomogeneous boundary conditions on
the nonlinear equations (\ref{eq:bcstrm}a,b), making $V_\theta$ proportional
to $Re$.  In order to make this dependence on $Re$ explicit, we will scale
$V_\theta$ (but not $V_{r,z}$ or $\Psi$) by $Re$:
\begin{equation}
V_\theta=Re \,U_\theta
\end{equation}
This scaling can be used even when $Re=0$ since $V_\theta$ is 
proportional to $Re$.
We then rewrite \eqref{eq:eqstrmlin} in a more compact matrix form as follows:
\begin{eqnarray}
\sigma
\left(\begin{array}{ccc} D^2 & 0 & 0 \\ 0 & I & 0 \\ 0 & 0 & I \end{array}\right)
\left(\begin{array}{c}\psi \\ \Theta \\ v_\theta \end{array}\right) 
=\mathcal{L}
\left(\begin{array}{c}\psi \\ \Theta \\ v_\theta \end{array}\right) \nonumber\\
=\left(\begin{array}{cc|c} \ml_{\psi\psi} 
& \ml_{\psi\Theta} & Re \,\ml_{\psi v_\theta} \\
\ml_{\Theta \psi} & \ml_{\Theta\Theta} &  0 \\\hline
Re\, \ml_{v_\theta \psi} & 0 & \ml_{v_\theta v_\theta} \\
\end{array}\right)
\left(\begin{array}{c}\psi \\ \Theta \\\hline v_\theta \end{array}\right)
\label{eq:matrixform}
\end{eqnarray}
The operators $\ml_{\psi\psi},\ldots$ which comprise $\ml$, and which depend
on $Re$ implicitly (and weakly) via the steady state, are listed
below.  The boxes surround terms which are non-zero when $Re=0$ and 
the steady state is the conductive solution $\bV=0$, $T=1-z$.
\begin{subequations}\begin{eqnarray}
\ml_{\psi\psi} \psi &=&
-\frac{1}{r}\left((\pd_z\psi)\,\pd_- -(\pd_r\psi)\,\pd_z\right)\,D^2\Psi \nonumber\\
&&-\left(V_r\pd_-+V_z\pd_z \right)\,D^2\psi 
+\boxed{\left(\lap-\frac{1}{r^2}\right)D^2\psi}\nonumber\\
\\
\ml_{\psi\Theta}\Theta&=&\boxed{\frac{Ra}{Pr}\pd_r\Theta}\\
\ml_{\psi v_\theta}v_\theta &=& \frac{2}{r}\pd_z (U_\theta v_\theta)\\
\ml_{\Theta\psi} \psi &=& 
-\frac{1}{r}(\pd_z\psi)\,\pd_rT +\boxed{\frac{1}{r}(\pd_r\psi)\,\pd_z\,T}\\
\ml_{v_\theta \psi} \psi &=& 
-\frac{1}{r}\left((\pd_z\psi)\,\pd_+ -(\pd_r\psi)\,\pd_z \right)\,U_\theta\\
\ml_{\Theta\Theta} \Theta &=& 
- (V_r\pd_r+V_z\pd_z)\, \Theta + \boxed{\frac{1}{Pr} \lap \Theta} \\
\ml_{v_\theta v_\theta} v_\theta &=& 
-\left(V_r\pd_++V_z\pd_z\right)\,v_\theta
+ \boxed{\left(\lap - \frac{1}{r^2}\right) v_\theta} 
\end{eqnarray}
\label{eq:subops}
\end{subequations}

Figure \ref{fig:arrange_eigs} shows the real part of the leading eigenvalues
(in units of the inverse viscous diffusive time) as a function of $Ra$ for
several fixed values of $Re$, while figure \ref{fig:evecs} shows some of the
corresponding leading eigenvectors for $Re=0$ and for $Re=96$, both for
$Ra=10\,000$.  Figure \ref{fig:arrange_eigs} shows several eigenvalues which
are independent of $Ra$ when $Re=0$.  The origin of these is well known and
easily understood. For $Re=0$, equations
\eqref{eq:matrixform}-\eqref{eq:subops} show that the linear stability problem
becomes:
\begin{eqnarray}
\sigma
\left(\begin{array}{ccc} D^2 & 0 & 0 \\ 0 & I & 0 \\ 0 & 0 & I \end{array}\right)
\left(\begin{array}{c}\psi \\ \Theta \\ v_\theta \end{array}\right) 
&&
\nonumber\\
=\left(\begin{array}{cc|c} \left(\lap-\frac{1}{r^2}\right)D^2 & -\frac{Ra}{Pr}\pd_r
& 0 \\
-\frac{1}{r}\pd_r & \frac{1}{Pr} \lap &  0 \\\hline
0 & 0 & \left(\lap - \frac{1}{r^2}\right) \\
\end{array}\right)
\left(\begin{array}{c}\psi \\ \Theta \\\hline v_\theta \end{array}\right)
&&
\label{eq:matrixform_re0}
\end{eqnarray}
with the homogeneous boundary conditions (\ref{eq:bcslin}).
Purely thermal eigenvectors have $\bv=0$, leading to:
\begin{equation}
\left.\begin{array}{l}
0 = -\frac{Ra}{Pr}\pd_r \Theta \\
\sigma\Theta=\frac{1}{Pr}\lap \Theta \\
0=\Theta|_{z=0,1} = \pd_r \Theta|_{r=0,1}
\end{array}\right\} \Longrightarrow
\begin{array}{l}\Theta=\sin (k\pi z) \\\mbox{with } \sigma = -(k\pi)^2
\end{array}
\end{equation}
The first of these thermal eigenvalues is 
seen in figure \ref{fig:arrange_eigs} ($Re=0$) as the horizontal line
very close to its analytic value of $-\pi^2=-9.8696$. 
The corresponding thermal eigenvector is shown in figure \ref{fig:evecs}c.
Another set of solutions contains only azimuthal velocity:
with $\Theta=\psi=0$ and $v_\theta$ a solution to 
\begin{equation}
\left.\begin{array}{l}
\sigma v_\theta =\left(\lap -\frac{1}{r^2}\right)v_\theta \\
0 = v_\theta|_{z=0,1}=v_\theta|_{r=0,1}
\end{array}\right\}\Longrightarrow
\begin{array}{l}
v_\theta = \sin (k\pi z)\: J_1(r j_{1n}) \\
\mbox{with } \sigma=-((k\pi)^2 + j_{1n}^2)\end{array}
\end{equation}
where $J_1$ is the first Bessel function and $j_{1n}=3.8317$, $7.0156$, $\ldots$
is one of the zeros of $J_1$.  The first of these azimuthal velocity
eigenvalues is the horizontal line seen in figure \ref{fig:arrange_eigs}
($Re=0$), 
again very close to its
analytically computed value of $-(\pi^2+3.8317^2)=-24.551$.  The corresponding
azimuthal velocity eigenvector is shown in figure \ref{fig:evecs}d.

The other eigenvalues shown in figure \ref{fig:arrange_eigs} ($Re=0$) 
increase with $Ra$. The zero crossings of the two largest eigenvalues are 
associated with the pitchfork bifurcations discussed extensively 
in the previous section; these take place at $Ra=\RacRezero$ and $Ra=\RacRezeroPFtwo $.
Their associated eigenvectors are shown in figure \ref{fig:evecs}a,b, 
for $Ra=10\,000$, where their eigenvalues are 38.5 and 15.9,
and they are seen to contain one and two concentric radial rolls, 
respectively. 
The leading eigenvectors for higher $Re$ are similar.
The convective states in figures \ref{fig:convective_Re40} and 
\ref{fig:convective_Re90} do not resemble figure \ref{fig:evecs}a, 
even though they result from a pitchfork bifurcation 
involving this eigenvector. This is because the nonlinear steady states 
in figures \ref{fig:convective_Re40} and \ref{fig:convective_Re90} 
are a superposition of these eigenvectors and the basic flows 
shown in figures \ref{fig:conductive_Re40} and \ref{fig:conductive_Re90}.
The eigenvector in figure \ref{fig:evecs}e, with two vertically 
stacked rolls, is associated with eigenvalue $\sigma=-21.2$, 
between that of the thermal and azimuthal eigenvalues at $Re=10\,000$.

An overall feature of the eigenvalues for $Re=0$ that can be seen in figure
\ref{fig:arrange_eigs} is that they are all real and cross transversely.
Indeed, it is well-known that for Rayleigh-B\'enard convection (i.e.~$Re=0$),
all of the eigenvalues are real.  When $Re>0$, the eigenvalue crossings which
occur for $Re=0$ become complex conjugate pairs. (Because the real parts of
the eigenvalues are shown in figure \ref{fig:arrange_eigs}, complex conjugate
pairs appear as a fusion of two eigenvalue curves.)  As $Re$ increases, the
$Ra$ intervals over which the eigenvalues are complex widen.  
By $Re=96$, the
real eigenvalues responsible for the pitchfork bifurcations have merged into a
complex conjugate pair whose real part crosses zero at $Ra=11\,856$ at the Hopf
bifurcation point.
At $Re=95$ (not shown), the fusion of two real eigenvalues into a
complex conjugate pair and the zero-crossing occur simultaneously 
at $Ra=\RaTakens$: this is the Takens-Bogdanov codimension-two point.

Figure \ref{fig:evecs}f, g shows the real and imaginary part of the complex
conjugate pair which is the leading eigenvector for $Re=96$.  The fact that
this complex pair at $Re=96$ originates in the fusion of the leading real
eigenvectors at lower $Re$ is made strikingly clear when the streamfunctions
$\psi$ in figure \ref{fig:evecs}f, g are compared with those of figure
\ref{fig:evecs}a, b.  Note that the choice of the real and imaginary parts of
an eigenvector is arbitrary, since an eigenvector can be multiplied by any
complex number. The particular choice here is imposed by the normalization
$\mathcal{I}m(\psi(r=1/2,z=1/2))=0$.  The decomposition of the complex eigenvector
$\psi$ into two vectors containing one and two concentric radial rolls shows
that the limit cycle will involve competing radial structures.  As with the
steady states, the lack of resemblance between figures \ref{fig:evecs}f, g and
the limit cycle in figure \ref{fig:re110_visu} is due to the
fact that these are superpositions of the basic state and the eigenvectors.

We investigate the progression of the eigenvalues from real to complex as $Re$
increases by examining the matrix in \eqref{eq:matrixform}, which is block
diagonal for $Re=0$, as shown in \eqref{eq:matrixform_re0}.  Its eigenvalues
and eigenvectors thus consist of two sets: those of the thermal convection
problem (upper left submatrix) and those of the azimuthal problem (lower right
submatrix). The convective, thermal and azimuthal eigenvalues cross
transversely because \eqref{eq:matrixform_re0} has no off-diagonal terms and
because the eigenvalues within each of these sub-problems, each associated
with a different spatial structure, do not cross one another.

This behavior resembles that seen near the Takens-Bogdanov point which occurs
in binary fluid or thermosolutal convection.  In the binary/thermosolutal
case, the pair of eigenvectors which interact have different origins: one can
be viewed as arising primarily from thermal convection and the other from
solutal convection \cite{Tuck_01}.  In the Rayleigh-B\'enard/von K\'arm\'an
case studied here, the eigenvectors which become complex both arise from
thermal convection; the difference between them is their spatial structure,
shown in figure \ref{fig:evecs}a,b.

In the binary/thermosolutal case, the transverse crossings undergo two
different fates, depending on the sign of the separation parameter $S$, which
describes whether the thermal and solutal convection act in concert or in
opposition.  For positive $S$, the eigenvalue curves separate into two
hyperbolas, in what is called avoided crossings, and remain real.  For
negative $S$, the eigenvalues join in complex conjugate pairs, as in figure
\ref{fig:arrange_eigs}.  These two cases can be understood in terms of a
$2\times 2$ matrix, whose off-diagonal terms are of the same sign if $S$ is
positive and of opposite signs if $S$ is negative.  Here, we have not
attempted such an analysis, but we can conclude that the off-diagonal matrices
\begin{equation}
\left(\begin{array}{c}  Re \, \ml_{\psi v_\theta} \\ 0 \end{array}\right)
\qquad\mbox{and}\qquad
\left(\begin{array}{cc} Re \, \ml_{v_\theta \psi} & 0 \end{array}\right)
\end{equation}
are in some sense of opposite signs, since the coupling they cause between 
the convective and azimuthal velocity eigenvectors leads to
complex eigenvalues.  The coupling between the convective and the 
purely thermal eigenvectors must also be of this type.
In \cite{Tuck_01}, a calculation of
the sign of the coupling terms is presented for the binary/thermosolutal case, 
involving projecting onto the eigenvectors of the diagonal submatrices.

\section{Discussion}

We have shown that a transition analogous to the onset of convection occurs in
an axisymmetric cylindrical container subjected to vertical gradients in
temperature and azimuthal velocity, i.e. in Rayleigh-B\'enard/von K\'arm\'an
flow.  Differential rotation, which causes mixing via Ekman pumping, delays
the onset of convection.  The transition is a pitchfork bifurcation, leading
to steady convection, for $Re < 95$ and a Hopf bifurcation, leading to
oscillatory convection, for $Re > 95$.  Between these two types of
convection, over most of the $(0\leq Re \leq 120, 0\leq Ra \leq 30\,000)$ 
parameter space, the transition occurs
via a SNIPER bifurcation, in which the stable steady states meet a pair of
unstable steady states and mutually annihilate, leaving a limit cycle in their
wake. Over a small portion of the parameter space, the scenario is more
complicated and involves several global bifurcations.
(Section \ref{sec:thresholds} mentions some consequences of relaxing 
the imposition of axisymmetry.)

The linear stability analysis of the axisymmetric 
Rayleigh-B\'enard/von K\'arm\'an problem also shows interesting features.
We have traced the way in which differential rotation 
couples the eigenvalue branches, which are real for Rayleigh-B\'enard
convection, in such a way that they become complex.
The close resemblance between the leading real pair of eigenmodes 
for $Re<95$ and the leading complex pair for $Re>95$ supports the 
idea that a common basis of eigenvectors could be used to make 
the reduced model studied in the companion paper \cite{Maria_09} more 
economical. 

Since its discovery by Andronov and Leontovich \cite{Andronov}, the SNIPER
bifurcation has appeared in a number of ODE systems used to model chemical
reactions, notably in excitable media and the Belousov-Zhabotinsky reaction
\cite{Guckenheimer}, and population biology, for example predator-prey
systems.  In the hydrodynamic and PDE context, the SNIPER bifurcation was
observed in simulations of axisymmetric Rayleigh-B\'enard convection
\cite{Tuck_Bark_88,Tuck_Bark_89,LRM_06,RLM_08}. Convective states with different
radial wavelengths compete, much as occurs in the present
Rayleigh-B\'enard/von K\'arm\'an configuration. The closest analogy with the
bifurcation scenario we observe is found in the small-aspect-ratio study of
Taylor-Couette flow by \cite{Abshagen1,Abshagen2}, in which rotating
and modulated rotating waves play the role of our steady states and limit
cycles, respectively.

Recently, the SNIPER bifurcation has been the subject of renewed attention as
a possible explanation for the reversals of the earth's magnetic field. A
dynamo engendered by bulk fluid motion was first produced in a laboratory
experiment of the von K\'arm\'an flow in sodium (VKS), an electrically
conducting fluid \cite{Berhanu_07}. This VKS experiment shows reversals of the
polarity of the magnetic field which bear some similarity to that of the
terrestial field. An explanation involving a SNIPER bifurcation in a
low-dimensional dynamical system with noise has been put forward to explain
these reversals \cite{Ravelet_PRL_08,Petrelis_09}.  This manifestation of the
SNIPER bifurcation also provides some possible justifications for studying the
axisymmetric flow. In the VKS experiment, the mean of the highly turbulent
flow is axisymmetric, even though the instantaneous flow is not.  
%
A different mechanism for magnetic field reversals has been proposed
\cite{Stefani_05}, in which noise is added to a low-dimensional dynamical
system displaying a Takens-Bogdanov point.  The Takens-Bogdanov point,
separating a steady from a Hopf bifurcation point, also plays an essential
role in our system, as it does in the investigations of axisymmetric
Rayleigh-B\'enard convection by \cite{Siggers} and of Taylor-Couette flow by
\cite{Abshagen1,Abshagen2}.  It seems plausible that the TB point and
the hysteresis point terminating the SNIPER bifurcation curve form part of the
unfolding of a codimension-three point.

Our study was first intended to determine the effect of the von K\'arm\'an flow 
on the onset of Rayleigh-B\'enard convection.
Over the course of the investigation, we encountered a bifurcation scenario
leading to oscillations, which should and does occur quite generally, whenever
consecutive pitchfork bifurcations take place.  It should therefore be of
interest to the dynamical-systems community as well as to fluid dynamicists.


\begin{thebibliography}{99}

\bibitem[Charlson \& Sani(1971)]{ChaSan71}
{G.~Charlson \& R.~Sani}, 
{\sl On thermoconvective instability in a bounded cylindrical fluid layer,}
{Int.~Journal.~Heat Mass Transfer\/} {\bf 14}, 2157--60 (1971).

\bibitem{Behringer}
R.P. Behringer, H. Gao, J.N. Shaumeyer
{\sl Time dependence in Rayleigh-B\'enard Convection with a 
Variable Cylindrical Geometry,}
Phys.~Rev.~Lett. {\bf 50}, 1199--1202 (1983).

\bibitem[Buell \& Catton(1983)]{BueCat}
{J.~Buell \& I.~Catton}, 
{\sl The effect of wall conduction on the stability of a fluid in a 
right circular cylinder heated from below,}
{J.~Heat Transfer\/} {\bf 105}, 255 (1983).

\bibitem[Wanschura {et~al.\/}(1996)Wanschura, Kuhlmann \& Rath]{WanKuhRat}
{M.~Wanschura, H.~C.~Kuhlmann \& H.~J.~Rath}, 
{\sl Three-dimensional instability of axisymmetric buoyant convection 
in cylinders heated from below,}
{J.~Fluid Mech.\/} {\bf 326}, 399--415 (1996).

\bibitem[Touihri {et~al.\/}(1999)Touihri, Ben~Hadid \& Henry]{TouHadHen}
{R.~Touihri, H.~Ben~Hadid \& D.~Henry}, 
{\sl On the onset of convective instabilities in cylindrical cavities heated 
from below.~I.~Pure thermal case,}
{Phys.~Fluids\/} {\bf 11}, 2078--2088 (1999).

\bibitem{Boronska_JFM}
{K.~Boro\'nska \& L.S.~Tuckerman},
{\sl Standing and travelling waves in cylindrical Rayleigh-Benard convection,}
{J.~Fluid Mech.} {\bf 559}, 279--298 (2006).

\bibitem{Tuck_Bark_88} 
L.S. Tuckerman \& D. Barkley, {\sl Global bifurcation to travelling waves
in axisymmetric convection,} Phys Rev. Lett. {\bf 61}, 408--411 (1988).

\bibitem{Tuck_Bark_89} 
D. Barkley \& L.S. Tuckerman, {\sl Traveling waves in axisymmetric
convection: the role of sidewall conductivity,} Physica D {\bf 37}, 288--294
(1989).

\bibitem[Siggers(2003)]{Siggers}
{J.~H.~Siggers}, 
{\sl Dynamics of targets in low-Prandtl number convection,}
{J.~Fluid Mech.\/} {\bf 475}, 357--375 (2003).

\bibitem[Gelfgat, Bar-Yoseph \& Solan (1996a)]{GBYS_96} 
Y.A.~Gelfgat, P.Z.~Bar-Yoseph \& A. Solan,
{\sl Steady states and oscillatory instability
    of swirling flow in a cylinder with rotating top and bottom,} 
Phys. Fluids {\bf 8}, 2614--2625 (1996).

\bibitem[Lopez (1998)]{L_98}
J.M.~Lopez,
{\sl Characteristics of endwall and sidewall boundary layers in a 
rotating cylinder with a differentially rotating endwall,}
J.~Fluid Mech. {\bf 359}, 49--79 (1998).

\bibitem[Lopez et al. (2002)]{LHMKS_02}
J.M.~Lopez, J.E.~Hart, F.~Marques, S.~Kittelman \& J.~Shen,
{\sl Instability and mode interactions in a differentially--driven rotating
  cylinder,} 
J.~Fluid Mech. {\bf 462}, 383--409 (2002).

\bibitem{MGL_03}
F. Marques, A.Y. Gelfgat, J.M. Lopez,
{\sl Tangent double Hopf bifurcation in a differentially rotating cylinder 
flow,} Phys. Rev. E {\bf 68}, 016310 (2003).

\bibitem{Nore_03}
C. Nore, L.S. Tuckerman, O. Daube, \& S. Xin, 
{\sl The 1:2 mode interaction in exactly counter-rotating von 
K\'arm\'an swirling flow,} J. Fluid Mech. {\bf 477}, 51--88 (2003).

\bibitem{Nore_04}
C. Nore, M. Tartar, O. Daube \& L. S. Tuckerman,
{\sl Survey of instability thresholds of 
flow between exactly counter-rotating disks,}
J. Fluid Mech. {\bf 511}, 45--65 (2004).

\bibitem{Nore_LMW_06} 
C. Nore, L. Martin Witkowski, E. Foucault, J. P\'echeux,
  O. Daube et P. Le Qu\'er\'e, 
{\sl Competition between axisymmetric and
  three-dimensional patterns between exactly counter-rotating disks,} 
Phys. Fluids {\bf 18}, 054102 (2006).

\bibitem{Brachet}
S. Douady, Y. Couder, M.E. Brachet,
{\sl Direct observation of the intermittency of intense vorticity 
filaments in turbulence,}
Phys. Rev. Lett. {\bf 67}, 983--986 (1991).

\bibitem{Ravelet_04}
F. Ravelet, L. Mari\'e, A. Chiffaudel, F. Daviaud,
{\sl Multistability and memory effect in a highly turbulent flow:
experimental evidence for a global bifurcation,}
Phys. Rev. Lett. {\bf 93}, 164501 (2004).

\bibitem{Berhanu_07} M. Berhanu, R. Monchaux, S. Fauve, N. Mordant,
  F. P\'etr\'elis, A. Chiffaudel, F. Daviaud, B. Dubrulle, C. Gasquet,
  L. Mari\'e, and F. Ravelet, M. Bourgoin, Ph. Odier, M. Moulin, J.-F. Pinton,
  R. Volk, 
{\sl Magnetic field reversals in an experimental turbulent dynamo,} 
Europhys.~Lett.   {\bf 77}, 59001 (2007)

\bibitem{Ravelet_PRL_08}
F. Ravelet, M. Berhanu, R. Monchaux, S. Aumaitre, A. Chiffaudel, F. Daviaud, 
B. Dubrulle, M. Bourgoin, P. Odier, N. Plihon, J.F. Pinton, R. Volk, S. Fauve, 
N. Mordant, F. Petrelis,
{\sl Chaotic dynamos generated by a turbulent flow of liquid sodium,}
Phys.~Rev.~Lett. {\bf 101}, 074502 (2008).

\bibitem{Petrelis_09}
F. P\'etr\'elis, S. Fauve, E. Dormy, J.-P. Valet,
{\sl Simple mechanism for reversals of earth's magnetic field,}
Phys. Rev. Lett. {\bf 102}, 144503 (2009).

\bibitem{Nicolas}
X. Nicolas,
{\sl Bibliographical review on the Poiseuille-Rayleigh-B\'enard  flows:
the mixed convection flows in horizontal rectangular ducts heated from below,}
Int.~J.~Thermal Sci.~{\bf 41}, 961--1016 (2002).

\bibitem{CB_91}
R.M. Clever, F.H. Busse, 
{\sl Instabilities of longitudinal rolls in the presence of Poiseuille flow,}
J.~Fluid Mech.~{\bf 229}, 517--529 (1991).

\bibitem{CB_92}
R.M. Clever, F.H. Busse,
{\sl Three-dimensional convection in a horizontal fluid layer 
subjected to a constant shear,} J.~Fluid Mech.~{\bf 234}, 511--527 (1992)

\bibitem{Muller_92}
H.W. M\"uller, M. L\"ucke, M. Kamps,
{\sl Transversal convection patterns in horizontal shear flow}.
Phys.~Rev.~A {\bf 45}, 3714--3726 (1992).

\bibitem{Lepiller}
V. Lepiller, A. Goharzadeh, A. Prigent, I. Mutabazi,
{\sl Weak temperature gradient effect on the stability of the circular Couette
flow,} Eur.~Phys.~J.~B {\bf 61}, 445--455 (2008).

\bibitem{TaggWeidman}
R. Tagg, P.D. Weidman, 
{\sl Linear stability of radially-heated circular Couette flow with simulated
radial gravity,}
Z. Angew. Math. Phys. {\bf 58}, 431--456 (2007).

\bibitem{OwenRogers1}
J.M. Owen, R.H. Rogers,
{\sl Flow and Heat Transfer in Rotating-Disk Systems, vol. 1: Rotor-Stator
Systems,} Wiley, 1989.

\bibitem{Knobloch_98}
E. Knobloch,
{\sl Rotating convection: Recent developments,}
 Int. J. Eng. Sci. {\bf 36}, 1421--1450 (1998). 

\bibitem{LRM_06}
J.M. Lopez, A. Rubio, F. Marques,
{\sl Travelling circular waves in axisymmetric rotating convection,}
J.~Fluid Mech.~{\bf 569}, 331-348 (2006).

\bibitem{RLM_08}
A. Rubio, J.M. Lopez, F. Marques,
{\sl Modulated rotating convection: radially travelling concentric rolls,}
J.~Fluid Mech.~{\bf 608}, 357--378 (2008).

\bibitem{Soong}
C.Y. Soong,
{\sl Theoretical analysis for axisymmetric mixed convection
between rotating coaxial disks,}
Int. J. Heat Mass Transfer {\bf 39}, 1569--1583 (1996).

\bibitem{Hill_Ball}
R.W. Hill, K.S. Ball, 
{\sl Direct numerical simulations of turbulent forced convection between 
counter-rotating disks,}
Int.~J.~Heat Fluid Flow {\bf 20}, 208--221 (1999).

\bibitem{Maria_09}
M.C. Navarro, L. Martin Witkowski, L.S. Tuckerman, P. Le Qu{\'e}r{\'e},
{\sl Building a reduced model for non-linear dynamics in Rayleigh-B{\'e}nard 
convection with counter-rotating disks,} Phys.~Rev.~E {\bf 81}, 036323 (2010).

\bibitem{Patankar}
S.V. Patankar, 
{\sl Numerical Heat Transfer and Fluid Flow,} 
Hemisphere Publishing, 1980.

\bibitem{LMW_JFM_01}
L. Martin Witkowski, P. Marty, J.S. Walker,
{\sl Liquid-metal flow in a finite-length cylinder with a high-frequency 
rotating magnetic field,}
J. Fluid Mech. {\bf 436}, 131--143 (2001).

\bibitem{Doedel}
E. Doedel,
Lecture notes on Numerical Analysis of Nonlinear Equations,
\url{http://cmvl.cs.concordia.ca/publications.notes.ps.gz}.

\bibitem{Govaerts}
W.J.F. Govaerts,
{\sl Numerical methods for bifurcations of dynamical equilibria,}
SIAM, 2000.

\bibitem{Maria_07}
M.C. Navarro, A.M. Mancho, H. Herrero,
{\sl Instabilities in buoyant flows under localized heating,}
Chaos {\bf 17}, 023105 (2007).

\bibitem{Houchens}
B.C.~Houchens, L.~Martin Witkowski, J.S.~Walker,
{\sl Rayleigh-B\'enard instability in a vertical cylinder with 
a vertical magnetic field,} J.~Fluid Mech.~{\bf 469}, 189--207 (2002). 

\bibitem{Andronov}
A. Andronov, E. Leontovich, 
{\sl Some cases of the dependence of the limit cycles upon parameters,}
Uchen. Zap. Gork. Univ. {\bf 6}, 3--24 (1939).

\bibitem{Kuznetsov}
Y. Kuznetsov,
{\sl Elements of Applied Bifurcation Theory,} Springer, 1998.

\bibitem{Arnold}
V. Arnold, 
{\sl Geometrical Methods in the Theory of Ordinary Differential Equations,} 
Springer, 1982.

\bibitem{Abshagen1}
J. Abshagen, J.M. Lopez, F. Marques \& G. Pfister,
{\sl Symmetry breaking via global bifurcations of modulated rotating waves in
hydrodynamics}, Phys. Rev. Lett. {\bf 94}, 074501 (2005).

\bibitem{Abshagen2}
J. Abshagen, J.M. Lopez, F. Marques \& G. Pfister,
{\sl Mode competition of rotating waves in reflection-symmetric Taylor-Couette
flow}, J. Fluid Mech. {\bf 540}, 269--299  (2005).


\bibitem{Tuck_01}
L.S. Tuckerman,
{\sl Thermosolutal and binary fluid convection as a $2\times 2$ matrix 
problem,} Physica D {\bf 156}, 325--363 (2001).

\bibitem{Guckenheimer}
J. Guckenheimer,
{\sl Multiple bifurcation problems for chemical reactions,}
Physica D {\bf 20}, 1--20 (1986).

\bibitem{Stefani_05}
F. Stefani, G. Gerbeth, 
{\sl Asymmetric polarity reversals, bimodal field distribution 
and coherence resonance in a spherically symmetric mean-field dynamo model,}
Phys. Rev. Lett. {\bf 94}, 184506 (2005).

\end{thebibliography}
\end{document}